\documentclass[twocolumn]{aastex631}
\usepackage{amsmath}
\usepackage{amssymb}
\usepackage{bm}
\usepackage{bbm}

\usepackage{float}
\usepackage{graphicx} 
\usepackage{multirow}
\usepackage{makecell}
\usepackage{newtxtext,newtxmath}
\usepackage{placeins}

\def\Euclid{\mbox{{\textit{Euclid}}}}

\definecolor{amber}{rgb}{1.0, 0.49, 0.0}
\definecolor{darkgreen}{rgb}{0.09, 0.45, 0.27}

\begin{document}
\title{Denoising Diffusion Probabilistic Model for realistic and fast generated \textit{Euclid}-like data for \\weak lensing analysis}

\author[0000-0001-8450-7885]{Diana Scognamiglio}
\affiliation{Jet Propulsion Laboratory, California Institute of Technology, 4800,  Oak Grove Drive - Pasadena, CA 91109, USA}

\author[0000-0002-2838-2878]{Jake H. Lee}
\affiliation{Jet Propulsion Laboratory, California Institute of Technology, 4800,  Oak Grove Drive - Pasadena, CA 91109, USA}

\author[0000-0002-9378-3424]{Eric Huff}
\affiliation{Jet Propulsion Laboratory, California Institute of Technology, 4800,  Oak Grove Drive - Pasadena, CA 91109, USA}

%\collaboration{20}{(AAS Journals Data Editors)}

\author[0000-0003-0220-0009]{Sergi R. Hildebrandt}
\affiliation{Jet Propulsion Laboratory, California Institute of Technology, 4800,  Oak Grove Drive - Pasadena, CA 91109, USA}

\author[0000-0003-2226-5395]{Shoubaneh Hemmati}
\affiliation{Infrared Processing and Analysis Center, California Institute of Technology, Pasadena, CA 01109, USA}



\begin{abstract}
%161 words
\noindent
Understanding and mitigating measurement systematics in weak lensing (WL) analysis requires large datasets of realistic galaxies with diverse morphologies and colors. Missions like \textit{Euclid}, the \textit{Nancy Roman} Space Telescope, and \textit{Vera C. Rubin} Observatory’s Legacy
Survey of Space and Time will provide unprecedented statistical power and control over systematic uncertainties. Achieving the stringent shear measurement requirement of $|m| < 10^{-3}$ demands analyzing $10^{9}$ galaxies. Accurately modeling galaxy morphology is crucial, as it is shaped by complex astrophysical processes that are not yet fully understood. Subtle deviations in shape and structural parameters can introduce biases in shear calibration. The interplay between bulges, disks, star formation, and mergers contributes to morphological diversity, requiring simulations that faithfully reproduce these features to avoid systematics in shear measurements. Generating such a large and realistic dataset efficiently is feasible using advanced generative models like denoising diffusion probabilistic models (DDPMs). In this work, we extend \textit{Hubble} Space Telescope (HST) data across \textit{Euclid}'s broad optical band using CANDELS and develop a generative AI tool to produce realistic \textit{Euclid}-like galaxies while preserving morphological details. We validate our tool through visual inspection and quantitative analysis of galaxy parameters, demonstrating its capability to simulate realistic \textit{Euclid} galaxy images, which will address WL challenges and enhance calibration for current and future cosmological missions.
\end{abstract}



\section{Introduction} \label{sec:intro}
In the era of precision cosmology, generating accurate and realistic galaxy images is fundamental to expanding our comprehension of the Universe. Current and upcoming astronomical surveys, including \Euclid\footnote{\url{https://sci.esa.int/Euclid/}}  \citep{laureijs2011euclid, Mellier2024euclid}, the \textit{Nancy Grace Roman} Space Telescope\footnote{\url{https://roman.gsfc.nasa.gov/}} \citep{spergel2015widefield}, and the \textit{Vera C. Rubin} Observatory’s Legacy
Survey of Space and Time (\textit{Rubin}-LSST\footnote{\url{https://www.lsst.org/lsst}}, \citealt{Ivezic2019LSST}) will observe billions of galaxies to study the structure and evolution of the cosmos. Among these, the \Euclid\ mission is specifically designed to map the geometry of the dark Universe by measuring the shapes and redshifts of galaxies over an area of about 14,000 deg$^2$ of the sky. A key scientific objective of \Euclid\ is to perform weak gravitational lensing (WL, see
e.g. \citealt{BertSchn01} for a detailed introduction) analysis, a technique that uses the subtle distortion of galaxy shapes caused by intervening matter along the line of sight to trace the distribution of dark and baryonic matter.

For WL analysis to reach its full potential, the weak amplitude of galaxy shape distortions—constituting only 1\% of their intrinsic shapes—demands highly precise measurements to avoid systematic errors that could bias shear estimates and, consequently, cosmological parameter analysis. To meet \Euclid’s stringent requirements on the shear measurements, the total error budget on shear calibration must be tightly constrained to the level of $10^{-4}$ \citep{Cropper_2013}, necessitating robust calibration methods and realistic galaxy simulations that reflect the diversity and observational conditions expected from \Euclid.

A significant challenge lies in creating such datasets. Since current simulations struggle to capture the full complexity of observed galaxy morphologies, empirical catalogs from deep, space-based surveys serve as a foundation for generating simulated datasets. Among these, the Cosmic Assembly Near-infrared Deep Extragalactic Legacy Survey (CANDELS, \citealt{2011Grogin, Koekemoer2011}), obtained with the \textit{Hubble} Space Telescope (HST), is widely used but remains limited in scale, containing only $10^5$ galaxies.
Moreover, simulations based on these datasets often are not fully able to reproduce the complex morphologies and color distributions observed in real galaxies \citep{Mandelbaum_2018, MacCrann2022, Castander2024}. Furthermore, traditional simulation software, like \texttt{GalSim} \citep{Rowe2015}, while effective for small-scale studies, become computationally demanding when scaled to the billions of galaxies needed for WL calibration. These limitations could lead to a reduced ability to fully align with \Euclid's observational requirements for high-precision WL analyses.

With the rise of deep learning in computer vision for tasks like image generation and classification, applying these advancements to generate realistic galaxy datasets presents a compelling solution to these challenges. Previously, \citet{Spindler_2020}, \citet{Lanusse2021}, and \citet{Holzschuh2022} demonstrated how Variational Autoencoders (VAEs, \citealt{Kingma2013}) and Generative Adversarial Networks (GANs, \citealt{goodfellow2014GAN, cohen2022GAN}) can be applied to generate galaxy images with high-resolution training data. This approach has since also been used for forecasting galaxy morphologies in \textit{Euclid} \citep{Bretonnière2022}. GANs, in particular, have been explored for deblending galaxy images \citep{Hemmati_2022}, highlighting their potential for addressing observational challenges in deep surveys. Beyond generative models, convolutional neural networks (CNNs) have been explored as a promising method to reconstruct true galaxy morphologies for WL shear bias calibration in \textit{Euclid} \citep{Csizi2024}. Recently, Denoising Diffusion Probabilistic Models (DDPMs, \citealt{ho2020DDPM, dhariwal2021diffusion, Smith2022, lizarraga2024}) have emerged as a state-of-the-art generative model class, achieving high-fidelity image generation, surpassing the other approaches.

In this work, we use high-resolution HST data to generate \Euclid-like galaxies in the VIS band \citep{Cropper_2013}, carefully accounting for the telescope’s observational properties, including its point spread function (PSF), wavelength range, and depth. These reprocessed galaxies serve as the training sample for a generative deep learning tool based on a DDPM architecture \citep{ho2020DDPM, dhariwal2021diffusion}. 
Additionally, we create a separate validation sample to assess the tool’s ability to generate galaxies with realistic morphologies—both isolated (singlet) and blended systems—while ensuring that the statistical distributions of key morphological properties, such as size, signal-to-noise ratio (SNR), and shape, are accurately reproduced. These properties are known to significantly impact shear calibration \citep{FenechConti2017, Mandelbaum_2018, Kannawadi2019}. Ensuring distributional fidelity is critical for minimizing biases in WL analyses and constitutes a novel aspect of this work. While developed for \Euclid\ calibration, this AI-driven framework is broadly applicable to future WL studies and upcoming surveys, including those conducted by the \textit{Roman} Space Telescope and the \textit{Rubin} Observatory.

The paper is structured as follows: section \ref{sec:data} introduces the HST data utilized for simulating \Euclid-like galaxies. Section \ref{sec:sim} details the process of transforming HST observations to emulate \Euclid\ observations. In section \ref{sec:DDPM}, we describe the DDPM developed to generate realistic \Euclid-like galaxy postage stamps. Section \ref{sec:results} focuses on validating the generated galaxies both by visual inspection and by comparing the joint distribution of their parameters with those in the validation dataset. Finally, section \ref{sec:conclusion} summarizes our findings and explores the broader implications of this tool for WL studies and other astronomical surveys.

\section{Data: HST Observations} \label{sec:data}
The dataset used for this study is based on observations from HST, specifically utilizing images from the Advanced Camera for Surveys (ACS) and the Wide Field Camera 3 (WFC3) as part of the Cosmic Assembly Near-infrared Deep Extragalactic Legacy Survey (CANDELS, \citealt{2011Grogin, Koekemoer2011}). The CANDELS dataset provides a catalog of 250,000 galaxies with high-resolution imaging, making it an ideal source for creating simulated \Euclid-like data. The images have $0.06^{\prime\prime}$ pixel scale, $0.08^{\prime\prime}$ full width at half maximum (FWHM) resolution, and $\sim$ 28.5 ($5\sigma$) depth in the F606W and F814W filter.  These observations cover a wide range of wavelengths, and for our scope, we use ACS images in the filters F435W, F606W, F775W, F814W, and F850LP, whose transmission curves are shown in Fig. \ref{fig:trans_curve}.

We adopt the methodology described in \citep{Hemmati2022} to generate $64 \times 64$ pixel postage stamps centered on galaxies in the CANDELS GOODS-S field. Galaxies are selected from the publicly available CANDELS GOODS-S catalog (\citealt{Guo2013}) based on several criteria: F160W magnitude brighter than 25 mag, redshift range $0.1<z<5$, \texttt{CLASS\_STAR} $<0.95$, and FWHM $> 1$ pixel. This study aims to create a \textit{Euclid}-like sample for WL analysis; therefore, we apply these cuts to the photometric catalog to remove galaxies whose shape distortions cannot be measured accurately enough for WL (e.g., \citealt{Hemmati2019}). While the selection is intentionally broad to include a diverse array of galaxy morphologies, sizes, and blending conditions, if specific regions of the color-magnitude-size space are found to be difficult for WL analysis and shape measurement, those could be excluded from lensing analysis later. This ensures the dataset's diversity and realism are maintained in the simulated outputs.

\section{Generating \Euclid-like Galaxies} \label{sec:sim}
\begin{figure}
    \centering  \includegraphics[width=\columnwidth]{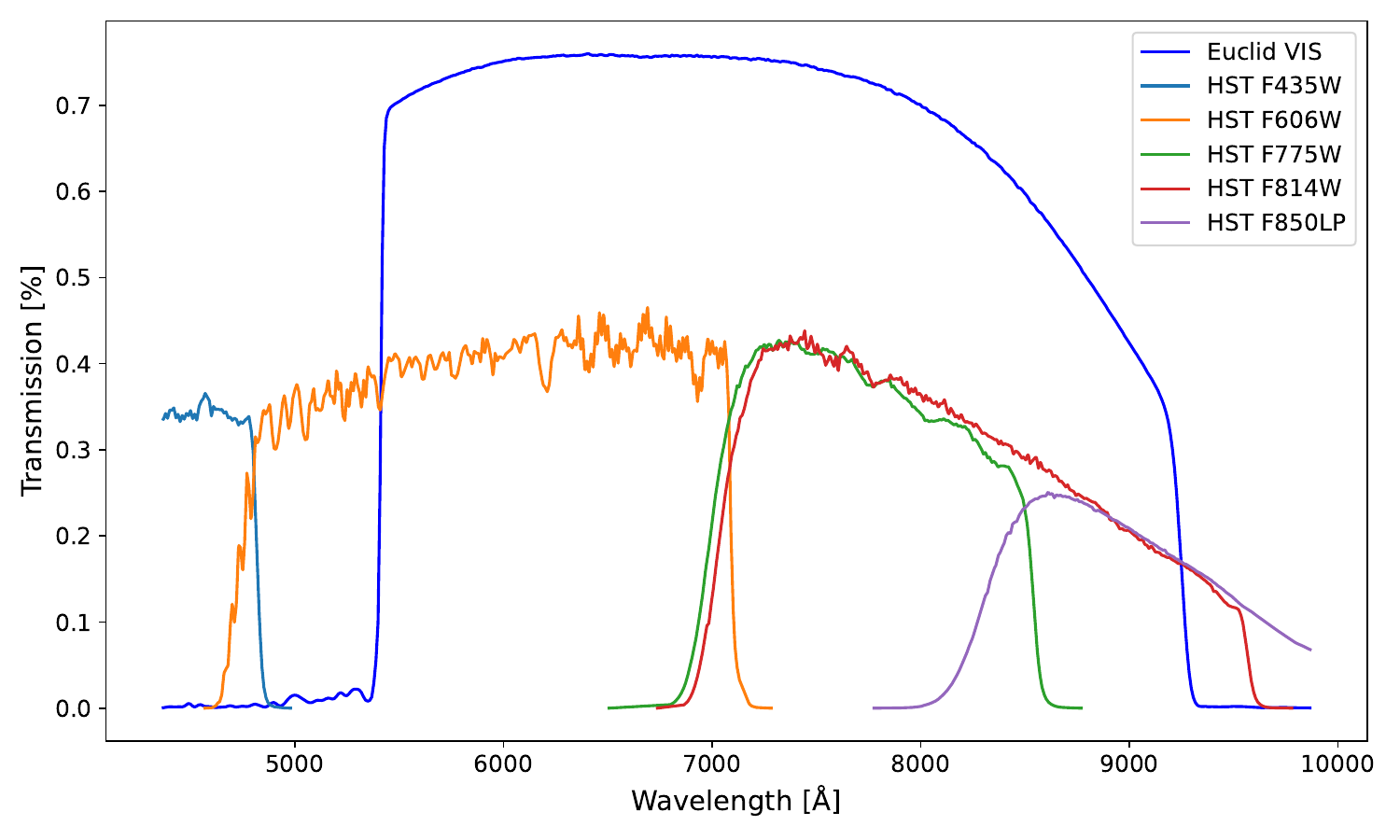}    
    \caption{Transmission curves of the \Euclid\ VIS, HST/ACS, and HST/WFC3 filters (F435W, F606W, F775W, F814W, and F850LP) used to simulate \Euclid-like galaxies based on HST galaxy images.}   \label{fig:trans_curve}
\end{figure}

In order to replicate the observational characteristics of the \Euclid\ VIS instrument and to generate \Euclid-like galaxy images from HST observations, we follow this procedure. The process begins with the deconvolution of the HST postage stamps by the HST PSF using \texttt{Galsim} \citep{Rowe2015}, isolating the intrinsic morphology of the galaxies. A Wiener filter is then applied with a $5 \times 5$ neighborhood to suppress noise while preserving fine structural details. This adaptive filter estimates the local noise variance within each neighborhood and adjusts the level of smoothing accordingly. By reducing high-frequency noise while maintaining the underlying morphological features, this step ensures a more stable deconvolution process.
The deconvolved images are subsequently convolved with a synthetic PSF designed to approximate the \Euclid\ VIS instrument’s PSF. This PSF is generated in \texttt{GalSim} as a diffraction-limited function with a 1.2-meter aperture and a central obscuration of $\approx 0.34$ \citep{Cropper_2013}, incorporating the effects of the telescope's secondary mirror.

Unlike empirical PSF models, centrally obscured diffraction patterns have well-defined analytic properties, making them particularly useful for WL analysis. Their mathematical simplicity facilitates the use of the generated images for WL shear bias calibration by streamlining the process of deconvolving the image by the synthetic PSF, applying a known shear, and reconvolving the resulting image with a realistic, spatially varying \Euclid\ PSF. This approach ensures that the intrinsic morphological properties of galaxies are preserved with minimal bias while maintaining control over observational effects.

After the convolution by the PSF, the images are then resampled from the HST ACS resolution of $0.06^{\prime\prime}$ pixel$^{-1}$ to \Euclid's $0.1^{\prime\prime}$ pixel$^{-1}$ to match the instrument's detector resolution. Spectral weighting is applied to account for differences in transmission properties between the HST filters (F435W, F606W, F775W, F814W, and F850LP) and the broad \Euclid\ VIS band. Transmission curves for each filter, shown in Fig. \ref{fig:trans_curve}, are used to compute weights that transform HST images into \Euclid-like observations, ensuring spectral consistency with the \Euclid\ VIS detector. To ensure the quality of the dataset, missing data in specific bands are excluded. We not that we do not explicitly model the wavelength dependence of the \Euclid\ PSF. Addressing this effect is essential for precision WL calibration but, we defer the inclusion of PSF chromaticity to follow-up work focused on shear measurement.

The resulting 30,588 \Euclid-like galaxy images are saved in HDF5 format. This dataset is split into 20,202 images for training and the remainder for validation. The training set is used to train the DDPM, which is utilized to generate new \Euclid-like observations (see Sect.~\ref{sec:DDPM}). The independent validation set, comprising 10,416 galaxy postage stamps selected without overlap with the training set but drawn from the same underlying distribution, is reserved for comparison with the generated \Euclid-like galaxies in Sect.~\ref{sec:results}.

% We train the DDPM models on two different datasets across xx CPUs, corresponding to $\sim$ xx hours wall time per model. We fill all available RAM and set the batch size to xx. ..

% \subsection{Conditional Generative Adversarial Network} \label{subsec:cGAN}
% Generative Adversarial Networks (GANs \diana{add ref}) have revolutionized the field of astronomical data analysis by providing innovative methods for generating synthetic datasets that closely resemble real observations. 

% At their core, GANs consist of a generator that creates artificial data samples, such as galaxy images or cosmic structures, and a discriminator that distinguishes these synthetic samples from actual observational data, in a dynamic, adversarial training process. This interaction results in highly realistic synthetic data, which can be used to augment limited observational datasets, enhance image resolution, or simulate rare astronomical events. Building on this framework, conditional Generative Adversarial Networks (cGANs\diana{add ref}) incorporate additional conditioning information, such as specific galaxy properties or redshift values, to guide the data generation process. This advancement allows allows cGANs to produce not only realistic but also tailored images that meets specific scientific criteria, significantly enhancing their utility for tasks such as image-to-image translation in astronomical surveys and the generation of high-fidelity simulations for weak lensing studies.

\section{Generative Model for \Euclid-like Simulations}
\label{sec:DDPM}
Denoising Diffusion Probabilistic Models (DDPMs, \citealt{ho2020DDPM, dhariwal2021diffusion}) are a class of deep generative models that generate diverse, high-resolution images. These models rely on a Markov chain framework to iteratively transform random noise into complex and realistic outputs. 

\begin{figure}[h!]
    \centering  \includegraphics[width=\columnwidth]{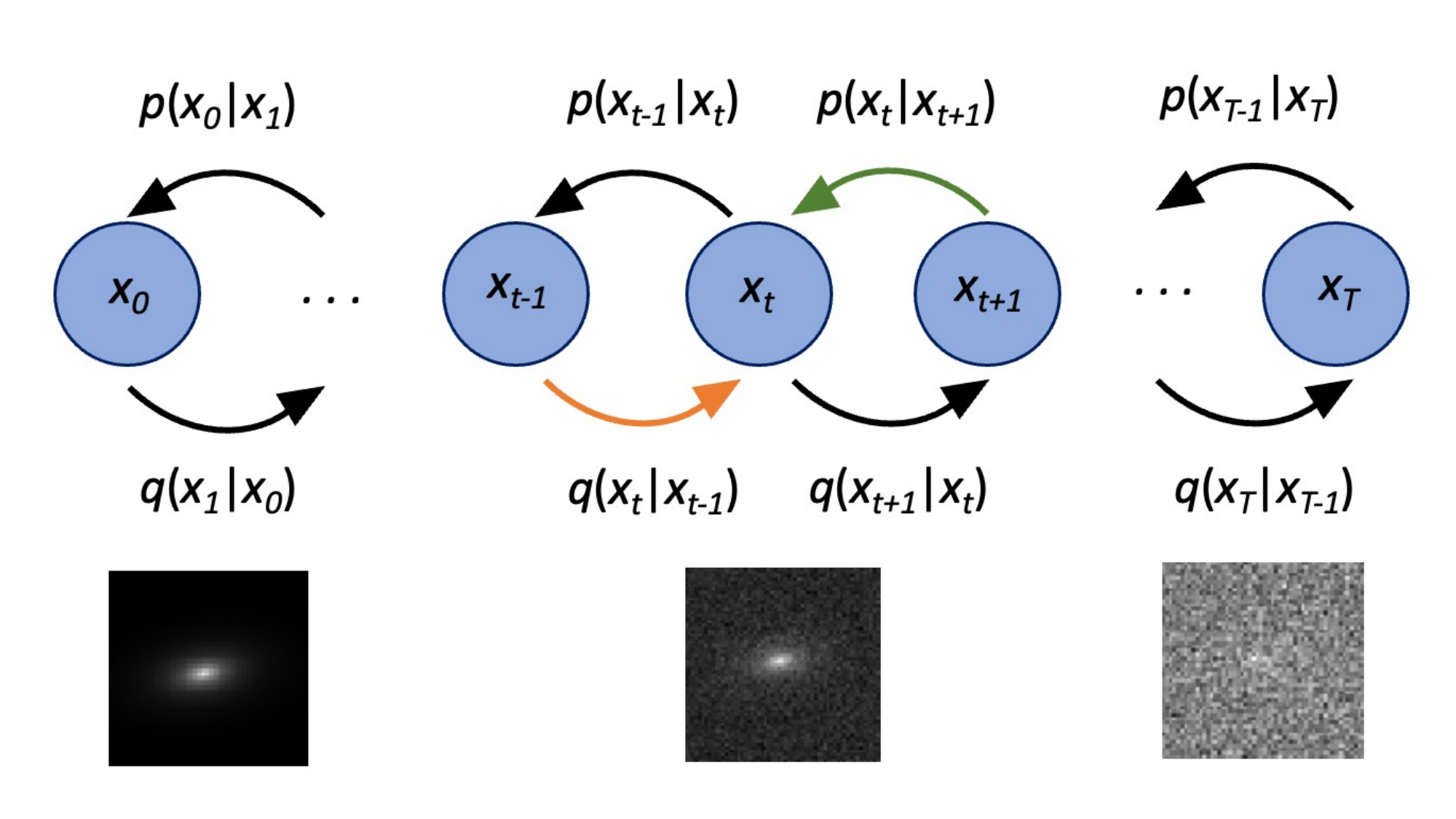}    
    \caption{Illustration of the Denoising Diffusion Probabilistic Model (DDPM) architecture. The three example images show the degradation and recovery of a galaxy through the iterative process.}   
    \label{fig:ddpm}
\end{figure}

The DDPM architecture, illustrated in Fig.~\ref{fig:ddpm}, consists of a forward diffusion process and a reverse denoising process. 

In the forward diffusion process, which represents the training phase, given clean \Euclid-like image $\mathbf{x}_{0}$ sampled from the data distribution $q(\mathbf{x})$ of the training dataset, small amounts of Gaussian noise are added to the data over $T$ steps, producing a sequence of noisy images, $\mathbf{x}_{t}$ with distribution $q(\mathbf{x}_{t}
|\mathbf{x}_{t-1})$. This process can be formulated as
\begin{equation}
    q(\textbf{x}_{t}|\textbf{x}_{\textbf{t}-1})=\mathcal{N}(\textbf{x}_{t};\mathbf{\mu}_{t}=\sqrt{1-\beta_{t}}\textbf{x}_{t-1},\mathbf{\Sigma}_{t}=\beta_{t}\textbf{I})\,,
\end{equation}
where $\bm{\mu}_{t}$ and $\mathbf{\Sigma}_{t}$ are the mean and the variance of the distribution, respectively, $\beta_{t}$ is the noise variance, equal for each dimension of the multi-dimensional space and, $\mathbf{I}$ is the identity matrix. So, we can go from the input data $\mathbf{x}_{0}$ to $\mathbf{x}_{T}$ in a tractable way. 

%The reverse process aims to recover the original data by denoising the noisy images step-by-step, and it corresponds to the phase where new \Euclid-like galaxies are generated. \textbf{We cannot easily estimate  $q(\mathbf{x}_t | \mathbf{x}_{t-1})$ because it needs to use the entire dataset, therefore we approximate it with a parameterized model $p_{\theta}$, with $\theta$ meaning that the parameters of the distribution of the reverse process are learned using a neural network.} 

The reverse process aims to recover the original data by denoising noisy images step-by-step, corresponding to the phase where new \Euclid-like galaxies are generated. We cannot directly compute the true reverse distribution $q(\mathbf{x}_{t-1} \vert \mathbf{x}_t) $ because it depends on the entire dataset's distribution $ q(\mathbf{x}_t) $, which is unknown. Instead, we approximate it with a parameterized model $ p_\theta(\mathbf{x}_{t-1} \vert \mathbf{x}_t) $, where $ \theta $ represents neural network parameters learned to match the ideal (but intractable) reverse process. This network predicts denoising steps using only the current noisy input $ \mathbf{x}_t $. Since $q(\mathbf{x}_t | \mathbf{x}_{t-1})$ is a Gaussian, for small $\beta_{t}$, we can choose $p_{\theta}$ to be Gaussian and just parameterize the mean and variance as follows
\begin{equation}
    p_{\theta}(\mathbf{x}_{t-1}|\mathbf{x}_{t})=\mathcal{N}(x_{t-1};\bm{\mu}_{\theta}(\mathbf{x}_{t},t),\mathbf{\Sigma}_{\theta}(\mathbf{x}_{t},t)).
\end{equation}

We base our implementation of the reverse process model on the work presented by \cite{dhariwal2021diffusion}. The parametrized model $p_{\theta}$ is implemented as a modified U-Net \citep{ronneberger2015u} model, with input/output dimensions of $32\times32$, 32 base channels, 2 residual blocks per down-sample, and added attention layers. We set the diffusion length to $T=500$ steps with linear noise scheduling; shorter diffusion steps resulted in degradation of image fidelity, while larger model architectures led to no significant improvements.

Taking $p(\textbf{x}_{T}) \sim \mathcal{N}(\textbf{x}_{T};\textbf{0},\mathbbm{1})$, which serves as the starting point for the reverse generative chain, we can use $p_{\theta}$ to iteratively sample from $p_{\theta}(x_{t-1} | x_{t})$ for $t=T, T-1,..., 1$, ultimately recovering entirely novel images that are similar, but not identical to, those found in the training set. This will allow us to augment the input \Euclid-like dataset and then generate from them new and realistic \Euclid\ data.

The benefit of the DDPM compared to the GAN \citep{goodfellow2014GAN, cohen2022GAN} includes greater sample diversity and visual fidelity, while being less prone to mode collapse.  The probabilistic nature of a DDPM inherently promotes a diverse range of outputs; each step in the denoising process is conditioned on the previous one, facilitating the exploration of multiple modes within the data distribution.

The diffusion model takes 11.5 hours to train for 500,000 steps, ensuring the convergence of the training loss function, with a batch size of 128 on a single NVIDIA A100 GPU. Image generation takes 0.432 seconds per stamp on the same hardware, highlighting the computational efficiency of this method. While further compute optimization and fidelity improvement through state-of-the-art sampling methods and noise scheduling is possible \citep{song2021denoising, lu2022dpm, karras2022elucidating}, this study focuses on the validation of the generated images.

\section{Validation of Generated Data} \label{sec:results}
\begin{figure*}[h!]
    \centering
    \raisebox{-0.5cm}{\textbf{DDPM-generated \Euclid-like galaxies}}\\ % Adjust vertical spacing using raisebox
    \includegraphics[width=0.72\textwidth]{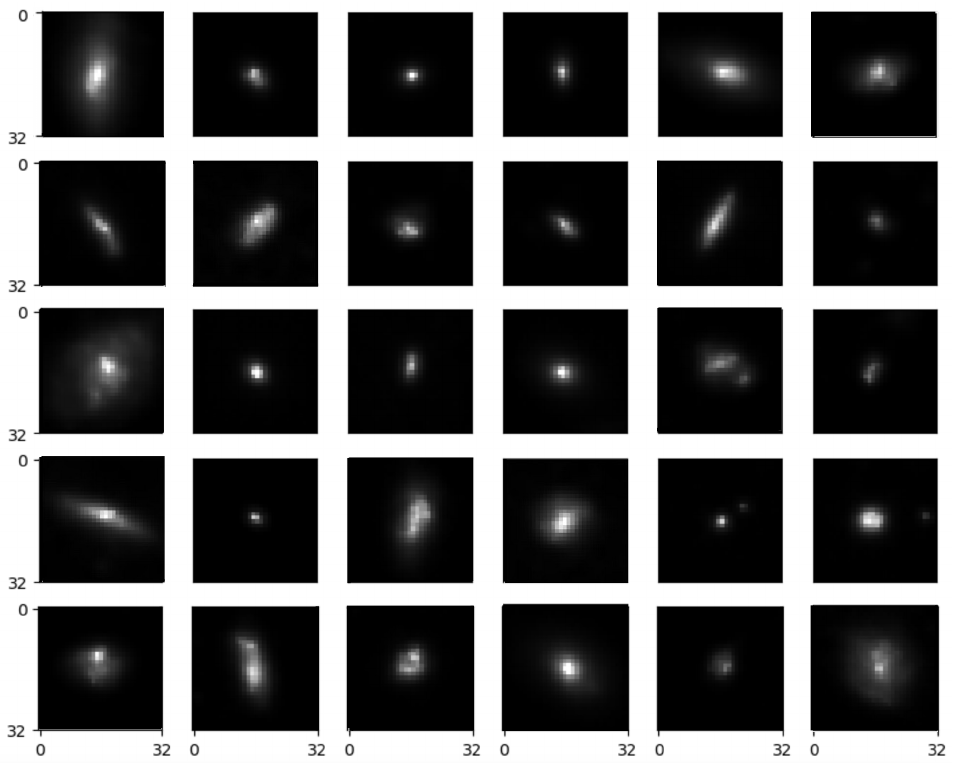}\\
    \vspace{-1.4cm}
    \raisebox{-1.6cm}{\textbf{\Euclid-like galaxies}}\\ 
    \includegraphics[width=0.72\textwidth]{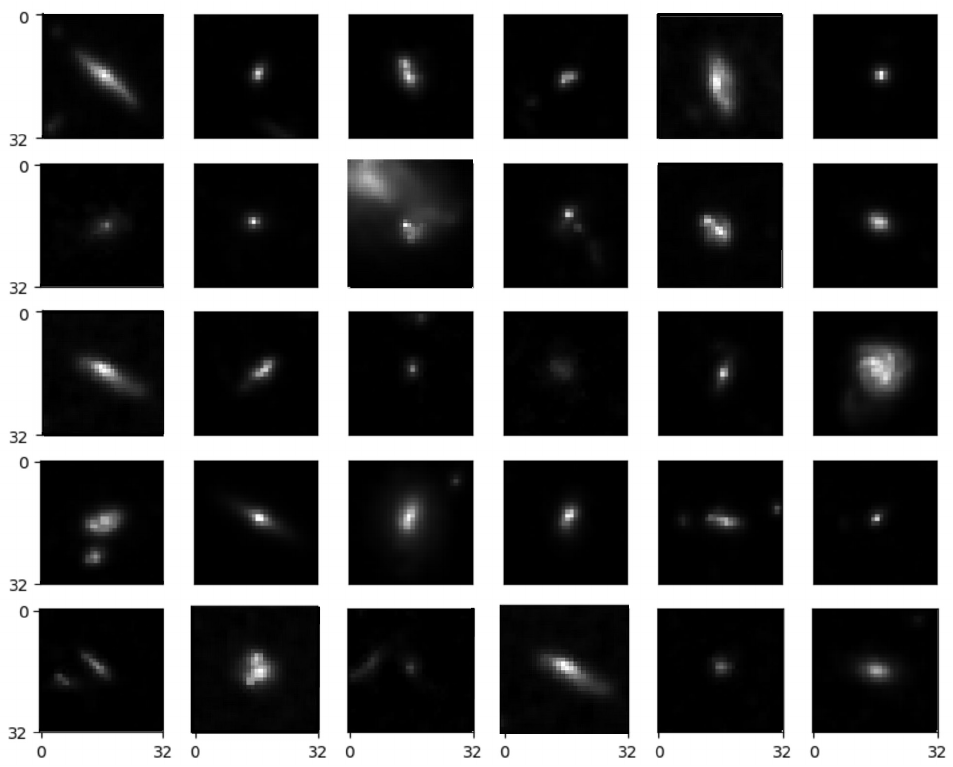}
    \caption{Comparison between generated galaxies (top) and validation galaxies (bottom) using the DDPM model. The generated images are produced using the DDPM model trained on \Euclid-like galaxy data, exhibiting realistic morphological features consistent with the training set. Each postage stamp is 32 $\times$ 32 pixels, corresponding to $3.2^{\prime\prime} \times 3.2^{\prime\prime}$ at the \Euclid\ pixel scale.}
    \label{fig:gal_comparison}
\end{figure*}
\begin{figure*}[t!]
    \centering
    \includegraphics[width=\textwidth]{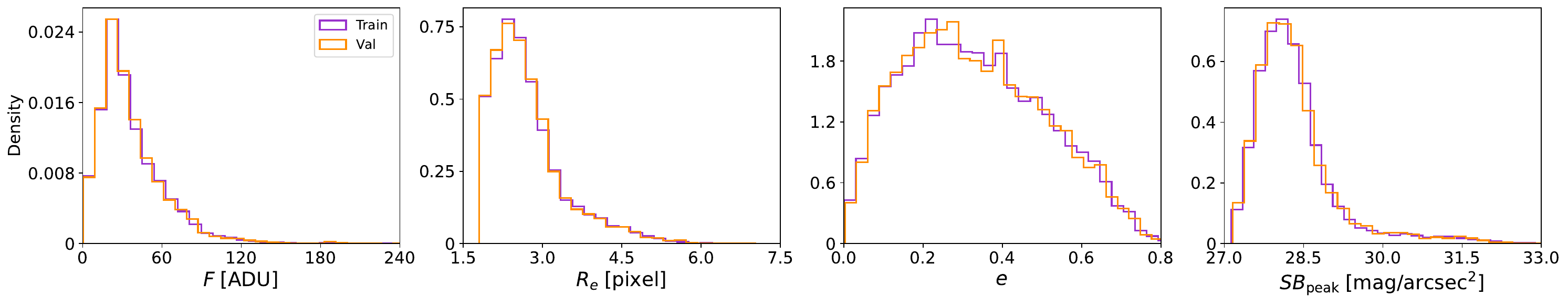}
    \caption{Distributions of Kron flux $F$ [ADU], effective radius $R_{\mathrm{e}}$ [pixel], ellipticity $e$, and peak surface brightness $SB_{\mathrm{peak}} [\mathrm{mag/arcsec^{2}]}$ for the training (Train) and validation (Val) samples.}
    \label{fig:hist_train_val}
\end{figure*}

Validating the realism of the DDPM-generated images is a crucial step to ensure their reliability for WL analysis and other scientific applications. This validation involves both visual and quantitative assessments. For the visual inspection, we generate $32 \times 32$ pixel postage stamp images, corresponding to $3.2^{\prime\prime} \times 3.2^{\prime\prime}$ at the \Euclid\ pixel scale. These images allow for direct comparisons between the generated data and the \Euclid-like validation data. Figure \ref{fig:gal_comparison} presents this comparison, with the top panel showing the DDPM-generated \Euclid-like galaxies, `Gen,’ and the bottom panel displaying the validation \Euclid-like galaxies, `Val.’ 
The visual similarity between the two datasets demonstrates the model’s ability to replicate diverse galaxy morphologies. Notably, it reproduces both singlets and blended cases, the latter comprising one-third of the sample and being more complex to model. The realistic appearance of generated images in both cases supports their suitability for scientific use.
In addition to visual validation, we perform a quantitative analysis to ensure the reliability of the generated galaxy properties. Using the \texttt{Photutils} library \citep{photutils}, we detect galaxies within each postage stamp image and measure four key properties: Kron flux ($F$), effective radius ($R_{\mathrm{e}}$), ellipticity ($e$), and peak surface brightness ($SB_{\mathrm{peak}}$). These quantities provide a simple yet effective characterization of flux, size, shape, and central concentration even for faint or poorly resolved sources, offering a robust way to evaluate whether the model preserves the statistical distribution of galaxy morphologies. The flux is directly measured, while $R_{\mathrm{e}}$ is computed as the mean of the semi-major and semi-minor axes; the ellipticity is defined as $e = 1 - \frac{\mathtt{semi\_minor\_sigma}}{\mathtt{semi\_major\_sigma}}$; and $SB_{\mathrm{peak}}$ is calculated from the maximum pixel intensity within each source segmentation, converted to physical units of mag arcsec$^{-2}$.

Before evaluating the generative performance, we verify the consistency between the training and validation datasets. As shown in Figure~\ref{fig:hist_train_val}, the distributions of $F$, $R_{\mathrm{e}}$, $e$, and $SB_{\mathrm{peak}}$ are in good agreement, confirming that both datasets are representative of the same underlying population.
The measurements of the galaxy parameters are then statistically compared across the validation and generated datasets to ensure consistency. For the comparison, we calculate the Kolmogorov-Smirnov (KS) statistic ($D^{\mathrm{KS}}$) over the entire dataset. To estimate the variability of the KS statistic and obtain a more robust understanding of the agreement between the distributions—particularly in the presence of potential outliers or non-uniform sampling— we perform a bootstrap analysis. In each of 5,000 iterations, we randomly draw 1,000 galaxies with replacement from each of the two datasets being compared (e.g., generated vs. validation), and compute the KS statistic between the two resampled distributions. We then report the range of KS-statistic values between the $1^{\mathrm{st}}$ and $99^{\mathrm{th}}$ percentiles as the bootstrap interval. This range provides a reference for expected variability due to sampling noise in the comparison. If the KS statistic for the generated dataset relative to the training set falls within this interval, the two distributions are statistically indistinguishable, given the statistical uncertainties imposed by the sample size.

The results from the statistical tests are summarized in Table \ref{tab:val_gen_comparison}.
\begin{table*}
\centering
\caption{Comparison of galaxy properties, including Kron flux ($F$) [ADU], effective radius ($R_{\mathrm{e}}$) [pixel], ellipticity ($e$), and peak surface brightness ($SB_{\mathrm{peak}}$) between the validation, `Val', and generated, `Gen' samples. The table presents the mean ($\mu$) and standard deviation ($\sigma$) for each parameter, the Kolmogorov–Smirnov (KS) statistic ($D^{\mathrm{KS}}$), and the bootstrap range for the KS statistic ($D_{[1-99]}^{\mathrm{bootstrap~ KS}}$).}
%, and the $p$-value from the $\chi^{2}$ test ($p^{\chi^{2}}$).} 
\renewcommand{\arraystretch}{1.3} % Adjust row height globally
\begin{tabular}{c|c|c|c|c}
\hline
\multicolumn{1}{c|}{Parameter} &
\multicolumn{1}{c|}{$\mu_{\mathrm{Val}} \pm \sigma_{\mathrm{Val}}$} &
\multicolumn{1}{c|}{$\mu_{\mathrm{Gen}} \pm \sigma_{\mathrm{Gen}}$} &
\multicolumn{1}{c|}{$D^{\mathrm{KS}}$} &
\multicolumn{1}{c}{$D_{\mathrm{[1-99]}}^{\mathrm{bootstrap~KS}}$} \\
%\multicolumn{1}{c}{$p^{\chi^{2}}$}  \\
\hline
$F$  [ADU]            & 36.6 $\pm$ 26.0 & 34.5 $\pm$ 22.7 & 0.033 & [0.025--0.046]  \\ %& $< 10^{-6}$ 
$R_{\mathrm{e}}$ [pixel] & 2.71 $\pm$ 0.70 & 2.82 $\pm$ 0.76 & 0.060  & [0.051--0.079] \\ %&  $< 10^{-6}$
$e$              & 0.33 $\pm$ 0.17 & 0.34 $\pm$ 0.18 & 0.017  & [0.014--0.035]  \\ %& 0.423
$SB_{\mathrm{peak}}$ [$\mathrm{mag/arsec^{-2}}$] & 28.35 $\pm$ 0.81 & 28.39 $\pm$ 0.91 & 0.045 & [0.000--0.629] \\

\hline
\end{tabular}
\label{tab:val_gen_comparison}
\end{table*}
The table includes the mean ($\mu$) and standard deviation ($\sigma$) for each parameter, along with the KS statistic ($D^{\mathrm{KS}}$) and its bootstrap range ($D_{\mathrm{[1-99]}}^{\mathrm{bootstrap~KS}}$). The mean and standard deviation values for the generative data closely match those of the validation data, as shown in the table. Moreover, the KS test statistics and their bootstrap-derived confidence intervals remain small for all parameters. These results demonstrate the generative model's capability to reproduce the underlying distributions of the validation dataset with high fidelity, particularly in terms of flux, size, and ellipticity.

\begin{figure*}[h!]
    \centering  
    \includegraphics[width=\textwidth]{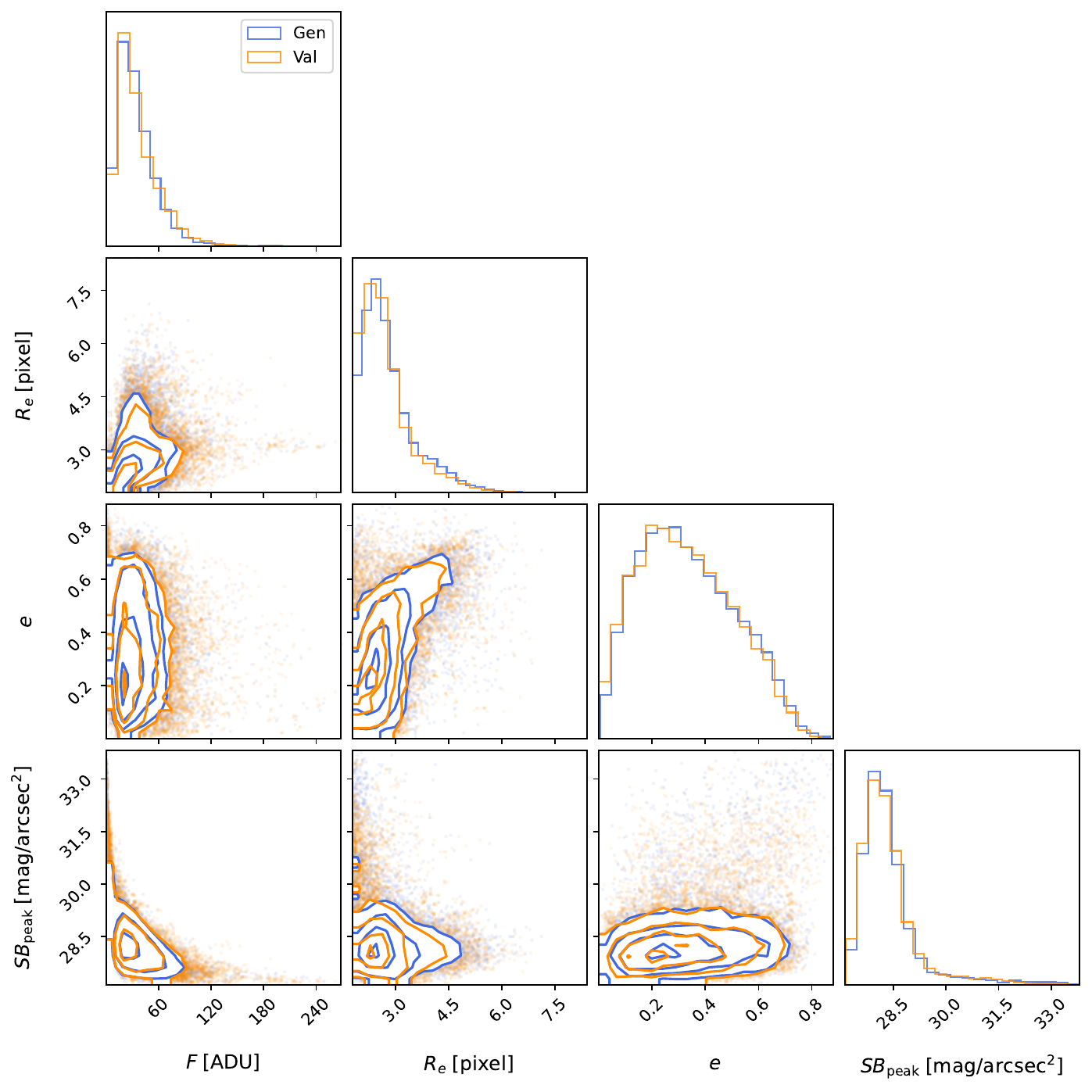}  
    \caption{Comparison of the joint normalized distributions of measured quantities: Kron flux $F$ [ADU], effective radius $R_{\mathrm{e}}$ [pixel], ellipticity $e$, and peak surface brightness $SB_{\mathrm{peak}} [\mathrm{mag/arsec^{2}}]$ between the generated (Gen) and \Euclid-like (Val) datasets. Contours in the scatter plots highlight the overlap of property correlations, confirming the robustness of the DDPM-generated dataset.}
    \label{fig:cornerplot}
\end{figure*}

Figure \ref{fig:cornerplot} illustrates the comparison between the validation and generated datasets, constituted of 10,416 and 10,712 galaxies postage stamps, in terms of galaxy parameters. The diagonal panels display normalized histograms of individual properties for the `Val' and `Gen' datasets, showing a good agreement in their distributions. For example, the flux distributions have means of $36.6 \pm 26.0$ ADU for the validation data and $34.5 \pm 22.7$ ADU for the generated data, with a KS statistic ($D^{\mathrm{KS}} = 0.033$) that lies within the bootstrap range [0.025--0.046]. Similarly, the ellipticity distributions exhibit excellent consistency, with means of $0.33 \pm 0.17$ and $0.34 \pm 0.18$, and the smallest KS statistic of $D^{\mathrm{KS}} = 0.017$. The size distributions are statistically consistent, with means of $2.82 \pm 0.76$ pixels for the generated sample and $2.71 \pm 0.70$ pixels for the validation sample. The KS statistic of $D^{\mathrm{KS}} = 0.060$ falls within the bootstrap range [0.051--0.079], confirming that the distributions are in agreement within statistical uncertainties. The peak surface brightness distribution is also well reproduced: $SB_{\mathrm{peak}} = 28.35 \pm 0.81$ mag arcsec$^{-2}$ for the validation sample and $28.39 \pm 0.91$ mag arcsec$^{-2}$ for the generated set, with a KS statistic of $D^{\mathrm{KS}} = 0.045$ that lies well within the bootstrap interval [0.000--0.629].

The off-diagonal panels in Fig.~\ref{fig:cornerplot} provide additional insights into the relationships between parameters. The 2D scatter plots and contour overlays demonstrate strong overlap, particularly in regions of higher density, indicating that the DDPM model accurately preserves the joint distributions of galaxy properties. This is particularly crucial for ensuring that the generated data faithfully reproduces the physical correlations observed in real datasets.

%Future work will focus on addressing minor discrepancies, particularly in blended cases, and evaluating the impact of these improvements on WL analyses and other applications.

%If we want to add also the subsets part
% The validation framework also includes jackknife tests, which involve splitting the training dataset into subsets based on median values of specific properties such as flux, size, and signal-to-noise ratio (SNR). This allows us to verify that the agreement between datasets holds across a range of parameter spaces. Corner plots in the figure show the relationships between pairs of properties (e.g., flux vs. size) for both datasets, along with contours that highlight the overlap in their distributions.

\section{Summary and Conclusion}
\label{sec:conclusion}
This study demonstrates that Denoising Diffusion Probabilistic Models (DDPMs, \citealt{ho2020DDPM}) can generate \Euclid-like galaxies with realistic morphologies, addressing key challenges in weak lensing (WL, \citealt{BertSchn01}) analysis and cosmological survey calibration. The generative model effectively learns and reproduces the morphological diversity of real galaxies, including both singlet and blended configurations, without requiring explicit input on these attributes. The generated galaxies successfully reproduce key properties—flux, effective radius, ellipticity, and peak surface brightness—which have the strongest impact on shear calibration. Their distributions closely match the validation data within $1\sigma$, providing a robust framework for augmenting datasets to meet the stringent precision requirements of shear measurements. These results demonstrate that the generative model accurately preserves not only the global morphological distributions but also key features related to galaxy central brightness.

The combination of visual inspection and statistical analyses, including Kolmogorov-Smirnov (KS) tests and bootstrap resampling, confirms the accuracy of the model in replicating the distributions of galaxy parameters. The DDPM-generated images offer a reliable representation of the \Euclid-like validation data. Furthermore, the model provides a computationally efficient solution, generating high-fidelity images in under a second per galaxy on standard GPU hardware.

The DDPM framework offers a powerful and scalable approach for generating synthetic galaxy images with realistic morphologies, enabling the creation of large datasets essential for WL calibration. These images facilitate the validation of shape measurement methods, improving accuracy by providing a diverse range of simulated galaxy properties. By producing datasets that closely match real observations, this work also aids in supporting pipeline development and calibration efforts for \Euclid\ \citep{Mellier2024euclid} and other upcoming surveys.

While developed specifically for \Euclid\ \citep{Mellier2024euclid}, this deep learning approach is broadly applicable to other cosmological missions, including the \textit{Nancy Grace Roman} Space Telescope \citep{spergel2015widefield} and the \textit{Vera C. Rubin} Observatory \citep{Ivezic2019LSST}. Unlike \Euclid, which observes in a single broad optical band, these surveys operate across multiple filters, requiring accurate multi-band representations of galaxy colors. Prior works have demonstrated the ability of generative diffusion models to learn and reproduce realistic color for computer vision tasks, suggesting their applicability for multi-wavelength tasks in the astrophysics domain. \citep{ho2020DDPM, Croitoru2023}
%The generative model’s ability to learn and reproduce realistic colors ensures its suitability for such applications, addressing the specific WL challenges of each mission. 
This study establishes a foundation for leveraging DDPMs to support precision cosmology, bridging the gap between data requirements and observational capabilities.

%% Also note that the akcnowlodgment environment does not support long amounts of text. If you have a lot of people and institutions to acknowledge, do not use this command. Instead, create a new 
\section*{Acknowledgements}
%\begin{acknowledgments}
The research was carried out at the Jet Propulsion Laboratory, California Institute of Technology, under a contract with the National
Aeronautics and Space Administration (80NM0018D0004), © 2024.
All rights reserved. In particular, this work was initiated with a JPL Data Science Pilot grant (E19001/DSI.100.24.05) under a program run by Daniel Crichton, Richard Doyle and, Miles Pellazar who provided early support and advice.
\vspace{0.5mm}

%\end{acknowledgments}


\vspace{5mm}
\facilities{All HST CANDELS data used in this paper can be found in MAST:\dataset[10.17909/T94S3X]{\doi{10.17909/T94S3X}}}


\software{Astropy \citep{2013A&A...558A..33A,2018AJ....156..123A}, Photutils \citep{photutils}
          }



%\appendix

\bibliography{biblio}{}

\begin{thebibliography}{}
\expandafter\ifx\csname natexlab\endcsname\relax\def\natexlab#1{#1}\fi
\providecommand{\url}[1]{\href{#1}{#1}}
\providecommand{\dodoi}[1]{doi:~\href{http://doi.org/#1}{\nolinkurl{#1}}}
\providecommand{\doeprint}[1]{\href{http://ascl.net/#1}{\nolinkurl{http://ascl.net/#1}}}
\providecommand{\doarXiv}[1]{\href{https://arxiv.org/abs/#1}{\nolinkurl{https://arxiv.org/abs/#1}}}

\bibitem[{{Astropy Collaboration} {et~al.}(2013){Astropy Collaboration}, {Robitaille}, {Tollerud}, {Greenfield}, {Droettboom}, {Bray}, {Aldcroft}, {Davis}, {Ginsburg}, {Price-Whelan}, {Kerzendorf}, {Conley}, {Crighton}, {Barbary}, {Muna}, {Ferguson}, {Grollier}, {Parikh}, {Nair}, {Unther}, {Deil}, {Woillez}, {Conseil}, {Kramer}, {Turner}, {Singer}, {Fox}, {Weaver}, {Zabalza}, {Edwards}, {Azalee Bostroem}, {Burke}, {Casey}, {Crawford}, {Dencheva}, {Ely}, {Jenness}, {Labrie}, {Lim}, {Pierfederici}, {Pontzen}, {Ptak}, {Refsdal}, {Servillat}, \& {Streicher}}]{2013A&A...558A..33A}
{Astropy Collaboration}, {Robitaille}, T.~P., {Tollerud}, E.~J., {et~al.} 2013, \aap, 558, A33, \dodoi{10.1051/0004-6361/201322068}

\bibitem[{{Astropy Collaboration} {et~al.}(2018){Astropy Collaboration}, {Price-Whelan}, {Sip{\H{o}}cz}, {G{\"u}nther}, {Lim}, {Crawford}, {Conseil}, {Shupe}, {Craig}, {Dencheva}, {Ginsburg}, {VanderPlas}, {Bradley}, {P{\'e}rez-Su{\'a}rez}, {de Val-Borro}, {Aldcroft}, {Cruz}, {Robitaille}, {Tollerud}, {Ardelean}, {Babej}, {Bach}, {Bachetti}, {Bakanov}, {Bamford}, {Barentsen}, {Barmby}, {Baumbach}, {Berry}, {Biscani}, {Boquien}, {Bostroem}, {Bouma}, {Brammer}, {Bray}, {Breytenbach}, {Buddelmeijer}, {Burke}, {Calderone}, {Cano Rodr{\'{\i}}guez}, {Cara}, {Cardoso}, {Cheedella}, {Copin}, {Corrales}, {Crichton}, {D'Avella}, {Deil}, {Depagne}, {Dietrich}, {Donath}, {Droettboom}, {Earl}, {Erben}, {Fabbro}, {Ferreira}, {Finethy}, {Fox}, {Garrison}, {Gibbons}, {Goldstein}, {Gommers}, {Greco}, {Greenfield}, {Groener}, {Grollier}, {Hagen}, {Hirst}, {Homeier}, {Horton}, {Hosseinzadeh}, {Hu}, {Hunkeler}, {Ivezi{\'c}}, {Jain}, {Jenness}, {Kanarek}, {Kendrew}, {Kern}, {Kerzendorf}, {Khvalko}, {King}, {Kirkby}, {Kulkarni},
  {Kumar}, {Lee}, {Lenz}, {Littlefair}, {Ma}, {Macleod}, {Mastropietro}, {McCully}, {Montagnac}, {Morris}, {Mueller}, {Mumford}, {Muna}, {Murphy}, {Nelson}, {Nguyen}, {Ninan}, {N{\"o}the}, {Ogaz}, {Oh}, {Parejko}, {Parley}, {Pascual}, {Patil}, {Patil}, {Plunkett}, {Prochaska}, {Rastogi}, {Reddy Janga}, {Sabater}, {Sakurikar}, {Seifert}, {Sherbert}, {Sherwood-Taylor}, {Shih}, {Sick}, {Silbiger}, {Singanamalla}, {Singer}, {Sladen}, {Sooley}, {Sornarajah}, {Streicher}, {Teuben}, {Thomas}, {Tremblay}, {Turner}, {Terr{\'o}n}, {van Kerkwijk}, {de la Vega}, {Watkins}, {Weaver}, {Whitmore}, {Woillez}, {Zabalza}, \& {Astropy Contributors}}]{2018AJ....156..123A}
{Astropy Collaboration}, {Price-Whelan}, A.~M., {Sip{\H{o}}cz}, B.~M., {et~al.} 2018, \aj, 156, 123, \dodoi{10.3847/1538-3881/aabc4f}

\bibitem[{{Bartelmann} \& {Schneider}(2001)}]{BertSchn01}
{Bartelmann}, M., \& {Schneider}, P. 2001, \physrep, 340, 291, \dodoi{10.1016/S0370-1573(00)00082-X}

\bibitem[{Bradley {et~al.}(2024)Bradley, Sip{\H o}cz, Robitaille, Tollerud, Vin{\'{\i}}cius, Deil, Barbary, Wilson, Busko, Donath, G{\"u}nther, Cara, Lim, Me{\ss}linger, Conseil, Burnett, Bostroem, Droettboom, Bray, Bratholm, Ginsburg, Jamieson, Barentsen, Craig, Morris, Perrin, Rathi, Pascual, \& Georgiev}]{photutils}
Bradley, L., Sip{\H o}cz, B., Robitaille, T., {et~al.} 2024, astropy/photutils: 2.0.2, 2.0.2,  Zenodo, \dodoi{10.5281/zenodo.13989456}

\bibitem[{Cohen \& Giryes(2022)}]{cohen2022GAN}
Cohen, G., \& Giryes, R. 2022, Generative Adversarial Networks.
\newblock \doarXiv{2203.00667}

\bibitem[{Croitoru {et~al.}(2023)Croitoru, Hondru, Ionescu, \& Shah}]{Croitoru2023}
Croitoru, F.-A., Hondru, V., Ionescu, R.~T., \& Shah, M. 2023, IEEE Transactions on Pattern Analysis and Machine Intelligence, 45, 10850, \dodoi{10.1109/TPAMI.2023.3261988}

\bibitem[{Cropper {et~al.}(2013)Cropper, Hoekstra, Kitching, Massey, Amiaux, Miller, Mellier, Rhodes, Rowe, Pires, Saxton, \& Scaramella}]{Cropper_2013}
Cropper, M., Hoekstra, H., Kitching, T., {et~al.} 2013, \mnras, 431, 3103, \dodoi{10.1093/mnras/stt384}

\bibitem[{Dhariwal \& Nichol(2021)}]{dhariwal2021diffusion}
Dhariwal, P., \& Nichol, A. 2021, in Advances in Neural Information Processing Systems, ed. M.~Ranzato, A.~Beygelzimer, Y.~Dauphin, P.~Liang, \& J.~W. Vaughan, Vol.~34 (Curran Associates, Inc.), 8780--8794.
\newblock \url{https://proceedings.neurips.cc/paper_files/paper/2021/file/49ad23d1ec9fa4bd8d77d02681df5cfa-Paper.pdf}

\bibitem[{Euclid Collaboration:~Bretonnière {et~al.}(2022)Euclid Collaboration:~Bretonnière, Huertas-Company, Boucaud, Lanusse, Jullo, Merlin, Tuccillo, Castellano, Brinchmann, Conselice, Dole, Cabanac, Courtois, Castander, Duc, Fosalba, Guinet, Kruk, Kuchner, Serrano, Soubrie, Tramacere, Wang, Amara, Auricchio, Bender, Bodendorf, Bonino, Branchini, Brau-Nogue, Brescia, Capobianco, Carbone, Carretero, Cavuoti, Cimatti, Cledassou, Congedo, Conversi, Copin, Corcione, Costille, Cropper, Da~Silva, Degaudenzi, Douspis, Dubath, Duncan, Dupac, Dusini, Farrens, Ferriol, Frailis, Franceschi, Fumana, Garilli, Gillard, Gillis, Giocoli, Grazian, Grupp, Haugan, Holmes, Hormuth, Hudelot, Jahnke, Kermiche, Kiessling, Kilbinger, Kitching, Kohley, Kümmel, Kunz, Kurki-Suonio, Ligori, Lilje, Lloro, Maiorano, Mansutti, Marggraf, Markovic, Marulli, Massey, Maurogordato, Melchior, Meneghetti, Meylan, Moresco, Morin, Moscardini, Munari, Nakajima, Niemi, Padilla, Paltani, Pasian, Pedersen, Pettorino, Pires, Poncet, Popa,
  Pozzetti, Raison, Rebolo, Rhodes, Roncarelli, Rossetti, Saglia, Schneider, Secroun, Seidel, Sirignano, Sirri, Stanco, Starck, Tallada-Crespí, Taylor, Tereno, Toledo-Moreo, Torradeflot, Valentijn, Valenziano, Wang, Welikala, Weller, Zamorani, Zoubian, Baldi, Bardelli, Camera, Farinelli, Medinaceli, Mei, Polenta, Romelli, Tenti, Vassallo, Zacchei, Zucca, Baccigalupi, Balaguera-Antolínez, Biviano, Borgani, Bozzo, Burigana, Cappi, Carvalho, Casas, Castignani, Colodro-Conde, Coupon, de~la Torre, Fabricius, Farina, Ferreira, Flose-Reimberg, Fotopoulou, Galeotta, Ganga, Garcia-Bellido, Gaztanaga, Gozaliasl, Hook, Joachimi, Kansal, Kashlinsky, Keihanen, Kirkpatrick, Lindholm, Mainetti, Maino, Maoli, Martinelli, Martinet, McCracken, Metcalf, Morgante, Morisset, Nightingale, Nucita, Patrizii, Potter, Renzi, Riccio, Sánchez, Sapone, Schirmer, Schultheis, Scottez, Sefusatti, Teyssier, Tutusaus, Valiviita, Viel, Whittaker, \& Knapen}]{Bretonnière2022}
Euclid Collaboration:~Bretonnière, H., Huertas-Company, M., Boucaud, A., {et~al.} 2022, A\&A, 657, A90, \dodoi{10.1051/0004-6361/202141393}

\bibitem[{{Euclid Collaboration:~Castander} {et~al.}(2024){Euclid Collaboration:~Castander}, {Fosalba}, {Stadel}, {Potter}, {Carretero}, {Tallada-Cresp{\'{\i}}}, {Pozzetti}, {Bolzonella}, {Mamon}, {Blot}, {Hoffmann}, {Huertas-Company}, {Monaco}, {Gonzalez}, {De Lucia}, {Scarlata}, {Breton}, {Linke}, {Viglione}, {Li}, {Zhai}, {Baghkhani}, {Pardede}, {Neissner}, {Teyssier}, {Crocce}, {Tutusaus}, {Miller}, {Congedo}, {Biviano}, {Hirschmann}, {Pezzotta}, {Aussel}, {Hoekstra}, {Kitching}, {Percival}, {Guzzo}, {Mellier}, {Oesch}, {Bowler}, {Bruton}, {Allevato}, {Gonzalez-Perez}, {Manera}, {Avila}, {Kov{\'a}cs}, {Aghanim}, {Altieri}, {Amara}, {Amendola}, {Andreon}, {Auricchio}, {Baldi}, {Balestra}, {Bardelli}, {Bender}, {Bodendorf}, {Bonino}, {Branchini}, {Brescia}, {Brinchmann}, {Camera}, {Capobianco}, {Carbone}, {Casas}, {Castellano}, {Cavuoti}, {Cimatti}, {Conselice}, {Conversi}, {Copin}, {Corcione}, {Courbin}, {Courtois}, {Da Silva}, {Degaudenzi}, {Di Giorgio}, {Dinis}, {Douspis}, {Dubath}, {Duncan}, {Dupac},
  {Dusini}, {Ealet}, {Farina}, {Farrens}, {Ferriol}, {Fotopoulou}, {Fourmanoit}, {Frailis}, {Franceschi}, {Franzetti}, {Galeotta}, {Gillard}, {Gillis}, {Giocoli}, {G{\'o}mez-Alvarez}, {Granett}, {Grazian}, {Grupp}, {Haugan}, {Holliman}, {Holmes}, {Hook}, {Hormuth}, {Hornstrup}, {Hudelot}, {Jahnke}, {Jhabvala}, {Joachimi}, {Keih{\"a}nen}, {Kermiche}, {Kiessling}, {Kilbinger}, {Kohley}, {Kubik}, {K{\"u}mmel}, {Kunz}, {Kurki-Suonio}, {Lahav}, {Laureijs}, {Le Mignant}, {Ligori}, {Lilje}, {Lindholm}, {Lloro}, {Maino}, {Maiorano}, {Mansutti}, {Marggraf}, {Markovic}, {Martinet}, {Marulli}, {Massey}, {Masters}, {Maurogordato}, {McCracken}, {Medinaceli}, {Mei}, {Melchior}, {Meneghetti}, {Merlin}, {Meylan}, {Mohr}, {Moresco}, {Moscardini}, {Munari}, {Nakajima}, {Nichol}, {Niemi}, {Padilla}, {Paech}, {Paltani}, {Pasian}, {Peacock}, {Pedersen}, {Pettorino}, {Pires}, {Polenta}, {Poncet}, {Popa}, {Raison}, {Rebolo}, {Renzi}, {Rhodes}, {Riccio}, {Romelli}, {Roncarelli}, {Rosset}, {Rossetti}, {Saglia}, {Sapone}, {Schirmer},
  {Schneider}, {Schrabback}, {Scodeggio}, {Secroun}, {Seidel}, {Serrano}, {Sirignano}, {Sirri}, {Stanco}, {Starck}, {Taylor}, {Teplitz}, {Tereno}, {Toledo-Moreo}, {Torradeflot}, {Tsyganov}, {Valenziano}, {Vassallo}, {Veropalumbo}, {Wang}, {Weller}, {Zacchei}, {Zamorani}, {Zerbi}, {Zoubian}, \& {Zucca}}]{Castander2024}
{Euclid Collaboration:~Castander}, F.~J., {Fosalba}, P., {Stadel}, J., {et~al.} 2024, arXiv e-prints, arXiv:2405.13495, \dodoi{10.48550/arXiv.2405.13495}

\bibitem[{{Euclid Collaboration: Csizi} {et~al.}(2024){Euclid Collaboration: Csizi}, {Schrabback}, {Grandis}, {Hoekstra}, {Jansen}, {Linke}, {Congedo}, {Taylor}, {Amara}, {Andreon}, {Baccigalupi}, {Baldi}, {Bardelli}, {Battaglia}, {Bender}, {Bodendorf}, {Bonino}, {Branchini}, {Brescia}, {Brinchmann}, {Camera}, {Capobianco}, {Carbone}, {Carretero}, {Casas}, {Castander}, {Castellano}, {Castignani}, {Cavuoti}, {Cimatti}, {Colodro-Conde}, {Conselice}, {Conversi}, {Copin}, {Courbin}, {Courtois}, {Cropper}, {Da Silva}, {Degaudenzi}, {De Lucia}, {Dinis}, {Douspis}, {Dubath}, {Dupac}, {Dusini}, {Farina}, {Farrens}, {Faustini}, {Ferriol}, {Fotopoulou}, {Frailis}, {Franceschi}, {Galeotta}, {Gillis}, {Giocoli}, {Grazian}, {Grupp}, {Guzzo}, {Haugan}, {Holmes}, {Hook}, {Hormuth}, {Hornstrup}, {Hudelot}, {Ili{\'c}}, {Jahnke}, {Jhabvala}, {Joachimi}, {Keih{\"a}nen}, {Kermiche}, {Kiessling}, {Kilbinger}, {Kubik}, {Kuijken}, {K{\"u}mmel}, {Kunz}, {Kurki-Suonio}, {Ligori}, {Lilje}, {Lindholm}, {Lloro}, {Maino}, {Maiorano},
  {Mansutti}, {Marcin}, {Marggraf}, {Markovic}, {Martinelli}, {Martinet}, {Marulli}, {Massey}, {Medinaceli}, {Mei}, {Melchior}, {Mellier}, {Meneghetti}, {Meylan}, {Moresco}, {Moscardini}, {Niemi}, {Padilla}, {Paltani}, {Pasian}, {Pedersen}, {Pettorino}, {Pires}, {Polenta}, {Poncet}, {Popa}, {Raison}, {Renzi}, {Rhodes}, {Riccio}, {Romelli}, {Roncarelli}, {Rossetti}, {Saglia}, {Sakr}, {S{\'a}nchez}, {Sartoris}, {Schneider}, {Secroun}, {Seidel}, {Serrano}, {Sirignano}, {Sirri}, {Stanco}, {Steinwagner}, {Tallada-Cresp{\'\i}}, {Tavagnacco}, {Teplitz}, {Tereno}, {Toledo-Moreo}, {Torradeflot}, {Tutusaus}, {Valentijn}, {Valenziano}, {Vassallo}, {Verdoes Kleijn}, {Veropalumbo}, {Wang}, {Weller}, {Zamorani}, {Zucca}, {Biviano}, {Bolzonella}, {Bozzo}, {Burigana}, {Calabrese}, {Di Ferdinando}, {Escartin Vigo}, {Farinelli}, {Gracia-Carpio}, {Matthew}, {Mauri}, {Pezzotta}, {P{\"o}ntinen}, {Scottez}, {Tenti}, {Viel}, {Wiesmann}, {Akrami}, {Allevato}, {Anselmi}, {Archidiacono}, {Atrio-Barandela}, {Ballardini}, {Blanchard},
  {Blot}, {Borgani}, {Bruton}, {Cabanac}, {Calabro}, {Ca{\~n}as-Herrera}, {Cappi}, {Caro}, {Carvalho}, {Castro}, {Chambers}, {Contarini}, {Cooray}, {Desprez}, {D{\'\i}az-S{\'a}nchez}, {Diaz}, {Di Domizio}, {Dole}, {Escoffier}, {Ferrari}, {Ferreira}, {Ferrero}, {Finoguenov}, {Fontana}, {Fornari}, {Gabarra}, {Ganga}, {Garc{\'\i}a-Bellido}, {Gasparetto}, {Gaztanaga}, \& {Giacomini}}]{Csizi2024}
{Euclid Collaboration: Csizi}, B., {Schrabback}, T., {Grandis}, S., {et~al.} 2024, arXiv e-prints, arXiv:2409.07528, \dodoi{10.48550/arXiv.2409.07528}

\bibitem[{Euclid Collaboration:~Mellier {et~al.}(2024)Euclid Collaboration:~Mellier, Abdurro'uf, Acevedo~Barroso, Ach\'ucarro, Adamek, Adam, Addison, Aghanim, Aguena, Ajani, Akrami, Al-Bahlawan, Alavi, Albuquerque, Alestas, Alguero, Allaoui, Allen, Allevato, Alonso-Tetilla, Altieri, Alvarez-Candal, Amara, Amendola, Amiaux, Andika, Andreon, Andrews, Angora, Angulo, Annibali, Anselmi, Anselmi, Arcari, Archidiacono, Aric\`o, Arnaud, Arnouts, Asgari, Asorey, Atayde, Atek, Atrio-Barandela, Aubert, Aubourg, Auphan, Auricchio, Aussel, Aussel, Avelino, Avgoustidis, Avila, Awan, Azzollini, Baccigalupi, Bachelet, Bacon, Baes, Bagley, Bahr-Kalus, Balaguera-Antolinez, Balbinot, Balcells, Baldi, Baldry, Balestra, Ballardini, Ballester, Balogh, Ba\~nados, Barbier, Bardelli, Barreiro, Barriere, Barros, Barthelemy, Bartolo, Basset, Battaglia, Battisti, Baugh, Baumont, Bazzanini, Beaulieu, Beckmann, Belikov, Bel, Bellagamba, Bella, Bellini, Benabed, Bender, Benevento, Bennett, Benson, Bergamini, Bermejo-Climent, Bernardeau,
  Bertacca, Berthe, Berthier, Bethermin, Beutler, Bevillon, Bhargava, Bhatawdekar, Bisigello, Biviano, Blake, Blanchard, Blazek, Blot, Bosco, Bodendorf, Boenke, B\"ohringer, Bolzonella, Bonchi, Bonici, Bonino, Bonino, Bonvin, Bon, Booth, Borgani, Borlaff, Borsato, Bosco, Bose, Botticella, Boucaud, Bouche, Boucher, Boutigny, Bouvard, Bouy, Bowler, Bozza, Bozzo, Branchini, Brau-Nogue, Brekke, Bremer, Brescia, Breton, Brinchmann, Brinckmann, Brockley-Blatt, Brodwin, Brouard, Brown, Bruton, Bucko, Buddelmeijer, Buenadicha, Buitrago, Burger, Burigana, Busillo, Busonero, Cabanac, Cabayol-Garcia, Cagliari, Caillat, Caillat, Calabrese, Calabro, Calderone, Calura, Camacho~Quevedo, Camera, Campos, Canas-Herrera, Candini, Cantiello, Capobianco, Cappellaro, Cappelluti, Cappi, Caputi, Cara, Carbone, Cardone, Carella, Carlberg, Carle, Carminati, Caro, Carrasco, Carretero, Carrilho, Carron~Duque, Carry, Carvalho, Carvalho, Casas, Casas, Casenove, Casey, Cassata, Castander, Castelao, Castellano, Castiblanco, Castignani,
  Castro, Cavet, Cavuoti, Chabaud, Chambers, Charles, Charlot, Chartab, Chary, Chaumeil, Cho, Chon, Ciancetta, Ciliegi, Cimatti, Cimino, Cioni, Claydon, Cleland, Cl\'ement, Clements, Clerc, Clesse, Codis, Cogato, Colbert, Cole, Coles, Collett, Collins, Colodro-Conde, Colombo, Combes, Conforti, Congedo, Conseil, Conselice, Contarini, Contini, Conversi, Cooray, Copin, Corasaniti, Corcho-Caballero, Corcione, Cordes, Corpace, Correnti, Costanzi, Costille, Courbin, Courcoult~Mifsud, Courtois, Cousinou, Covone, Cowell, Cragg, Cresci, Cristiani, Crocce, Cropper, E~Crouzet, Csizi, Cuby, Cucchetti, Cucciati, Cuillandre, Cunha, Cuozzo, Daddi, D'Addona, Dafonte, Dagoneau, Dalessandro, Dalton, D'Amico, Dannerbauer, Danto, Das, Da~Silva, da~Silva, Daste, Davies, Davini, de~Boer, Decarli, De~Caro, Degaudenzi, Degni, de~Jong, de~la Bella, de~la Torre, Delhaise, Delley, Delucchi, De~Lucia, Denniston, De~Paolis, De~Petris, Derosa, Desai, Desjacques, Despali, Desprez, De~Vicente-Albendea, Deville, Dias, D\'{\i}az-S\'anchez,
  Diaz, Di~Domizio, Diego, Di~Ferdinando, Di~Giorgio, Dimauro, Dinis, Dolag, Dolding, Dole, Dom\'{\i}nguez~S\'anchez, Dor\'e, Dournac, Douspis, Dreihahn, Droge, Dryer, Dubath, Duc, Ducret, Duffy, Dufresne, Duncan, Dupac, Duret, Durrer, Durret, Dusini, Ealet, Eggemeier, Eisenhardt, Elbaz, Elkhashab, Ellien, Endicott, Enia, Erben, Escartin~Vigo, Escoffier, Escudero~Sanz, Essert, Ettori, Ezziati, Fabbian, Fabricius, Fang, Farina, Farina, Farinelli, Farrens, Faustini, Feltre, Ferguson, Ferrando, Ferrari, Ferr\'e-Mateu, Ferreira, Ferreras, Ferrero, Ferriol, Ferruit, Filleul, Finelli, Finkelstein, Finoguenov, Fiorini, Flentge, Focardi, Fonseca, Fontana, Fontanot, Fornari, Fosalba, Fossati, Fotopoulou, Fouchez, Fourmanoit, Frailis, Fraix-Burnet, Franceschi, Franco, Franzetti, Freihoefer, Frittoli, Frugier, Frusciante, Fumagalli, Fumagalli, Fumana, Fu, Gabarra, Galeotta, Galluccio, Ganga, Gao, Garc\'{\i}a-Bellido, Garcia, Gardner, Garilli, Gaspar-Venancio, Gasparetto, Gautard, Gavazzi, Gaztanaga, Genolet,
  Genova~Santos, Gentile, George, Ghaffari, Giacomini, Gianotti, Gibb, Gillard, Gillis, Ginolfi, Giocoli, Girardi, Giri, Goh, G\'omez-Alvarez, Gonzalez, Gonzalez, Gonzalez, Gouyou~Beauchamps, Gozaliasl, Gracia-Carpio, Grandis, Granett, Granvik, Grazian, Gregorio, Grenet, Grillo, Grupp, Gruppioni, Gruppuso, Guerbuez, Guerrini, Guidi, Guillard, Gutierrez, Guttridge, Guzzo, Gwyn, Haapala, Haase, Haddow, Hailey, Hall, Hall, Hamaus, Haridasu, Harnois-D\'eraps, Harper, Hartley, Hasinger, Hassani, Hatch, Haugan, H{\"a}u{\ss}ler, Heavens, Heisenberg, Helmi, Helou, Hemmati, Henares, Herent, Hern\'andez-Monteagudo, Heuberger, Hewett, Heydenreich, Hildebrandt, Hirschmann, Hjorth, Hoar, Hoekstra, Holland, Holliman, Holmes, Hook, Horeau, Hormuth, Hornstrup, Hosseini, Hu, Hudelot, Hudson, Huertas-Company, Huff, Hughes, Humphrey, Hunt, Huynh, Ibata, Ichikawa, Iglesias-Groth, Ilbert, Ili\'c, Ingoglia, Iodice, Israel, Israelsson, Izzo, Jablonka, Jackson, Jacobson, Jafariyazani, Jahnke, Jansen, Jarvis, Jasche, Jauzac, Jeffrey,
  Jhabvala, Jimenez-Teja, Jimenez Mu\~noz, Joachimi, Johansson, Joudaki, Jullo, Kajava, Kang, Kannawadi, Kansal, Karagiannis, K\"archer, Kashlinsky, Kazandjian, Keck, Keih\"anen, Kerins, Kermiche, Khalil, Kiessling, Kiiveri, Kilbinger, Kim, King, Kirkpatrick, Kitching, Kluge, Knabenhans, Knapen, Knebe, Kneib, Kohley, Koopmans, Koskinen, Koulouridis, Kou, Kov\'acs, Kova\\vc\i\'c, Kowalczyk, Koyama, Kraljic, Krause, Kruk, Kubik, Kuchner, Kuijken, K\"ummel, Kunz, Kurki-Suonio, Lacasa, Lacey, La~Franca, Lagarde, Lahav, Laigle, La~Marca, La~Marle, Lamine, Lam, Lan{\c{c}}on, Landt, Langer, Lapi, Larcheveque, Larsen, Lattanzi, Laudisio, Laugier, Laureijs, Lavaux, Lawrenson, Lazanu, Lazeyras, Le~Boulc'h, Le~Brun, Le~Brun, Leclercq, Lee, Le~Graet, Legrand, Leirvik, Le~Jeune, Lembo, Le~Mignant, Lepinzan, Lepori, Lesci, Lesgourgues, Leuzzi, Levi, Liaudat, Libet, Liebing, Ligori, Lilje, Lin, Linde, Linder, Lindholm, Linke, Li, Liu, Lloro, Lobo, Lodieu, Lombardi, Lombriser, Lonare, Longo, L\'opez-Caniego, Lopez~Lopez,
  Alvarez, Loureiro, Loveday, Lusso, Macias-Perez, Maciaszek, Magliocchetti, Magnard, Magnier, Magro, Mahler, Mainetti, Maino, Maiorano, Maiorano, Malavasi, Mamon, Mancini, Mandelbaum, Manera, Manj\'on-Garc\'{\i}a, Mannucci, Mansutti, Manteiga~Outeiro, Maoli, Maraston, Marcin, Marcos-Arenal, Margalef-Bentabol, Marggraf, Marinucci, Marinucci, Markovic, Marleau, Marpaud, Martignac, Mart\'{\i}n-Fleitas, Martin-Moruno, Martin, Martinelli, Martinet, Martin, Martins, Marulli, Massari, Massey, Masters, Matarrese, Matsuoka, Matthew, Maughan, Mauri, Maurin, Maurogordato, McCarthy, McConnachie, McCracken, McDonald, McEwen, McPartland, Medinaceli, Mehta, Mei, Melchior, Melin, M\'enard, Mendes, Mendez-Abreu, Meneghetti, Mercurio, Merlin, Metcalf, Meylan, Migliaccio, Mignoli, Miller, Miluzio, Milvang-Jensen, Mimoso, Miquel, Miyatake, Mobasher, Mohr, Monaco, Mongui\'o, Montoro, Mora, Moradinezhad~Dizgah, Moresco, Moretti, Morgante, Morisset, Moriya, Morris, Mortlock, Moscardini, Mota, Moustakas, Moutard, M\"uller, Munari,
  Murphree, Murray, Murray, Musi, Nadathur, Nagam, Nagao, Naidoo, Nakajima, Nally, Natoli, Navarro-Alsina, Navarro~Girones, Neissner, Nersesian, Nesseris, Nguyen-Kim, Nicastro, Nichol, Nielbock, Niemi, Nieto, Nilsson, Noller, Norberg, Nourizonoz, Ntelis, Nucita, Nugent, Nunes, Nutma, Ocampo, Odier, Oesch, Oguri, Magalhaes~Oliveira, Onoue, Oosterbroek, Oppizzi, Ordenovic, Osato, Pacaud, Pace, Padilla, Paech, Pagano, Page, Palazzi, Paltani, Pamuk, Pandolfi, Paoletti, Paolillo, Papaderos, Pardede, Parimbelli, Parmar, Partmann, Pasian, Passalacqua, Paterson, Patrizii, Pattison, Paulino-Afonso, Paviot, Peacock, Pearce, Pedersen, Peel, Peletier, Pellejero~Ibanez, Pello, Penny, Percival, Perez-Garrido, Perotto, Pettorino, Pezzotta, Pezzuto, Philippon, Piersanti, Pietroni, Piga, Pilo, Pires, Pisani, Pizzella, Pizzuti, Plana, Polenta, Pollack, Poncet, P\"ontinen, Pool, Popa, Popa, Popp, Porciani, Porth, Potter, Poulain, Pourtsidou, Pozzetti, Prandoni, Pratt, Prezelus, Prieto, Pugno, Quai, Quilley, Racca, Raccanelli,
  R\'acz, Radinovi\'c, Radovich, Ragagnin, Ragnit, Raison, Ramos-Chernenko, Ranc, Raylet, Rebolo, Refregier, Reimberg, Reiprich, Renk, Renzi, Retre, Revaz, Reyl\'e, Reynolds, Rhodes, Ricci, Ricci, Riccio, Ricken, Rissanen, Risso, Rix, Robin, Rocca-Volmerange, Rocci, Rodenhuis, Rodighiero, Rodriguez~Monroy, Rollins, Romanello, Roman, Romelli, Romero-Gomez, Roncarelli, Rosati, Rosset, Rossetti, Roster, Rottgering, Rozas-Fern\'andez, Ruane, Rubino-Martin, Rudolph, Ruppin, Rusholme, Sacquegna, S\'aez-Casares, Saga, Saglia, Sahl\'en, Saifollahi, Sakr, Salvalaggio, Salvaterra, Salvati, Salvato, Salvignol, S\'anchez, Sanchez, Sanders, Sapone, Saponara, Sarpa, Sarron, Sartori, Sassolas, Sauniere, Sauvage, Sawicki, Scaramella, Scarlata, Scharr\'e, Schaye, Schewtschenko, Schindler, Schinnerer, Schirmer, Schmidt, Schmidt, Schmidt, Schneider, Schneider, Schneider, Sch\"oneberg, Schrabback, Schultheis, Schulz, Schwartz, Sciotti, Scodeggio, Scognamiglio, Scott, Scottez, Secroun, Sefusatti, Seidel, Seiffert, Sellentin,
  Selwood, Semboloni, Sereno, Serjeant, Serrano, Shankar, Sharples, Short, Shulevski, Shuntov, Sias, Sikkema, Silvestri, Simon, Sirignano, Sirri, Skottfelt, Slezak, Sluse, Smith, Smith, Smith, Smit, Soldano, Solheim, Sorce, Sorrenti, Soubrie, Spinoglio, Spurio~Mancini, Stadel, Stagnaro, Stanco, Stanford, Starck, Stassi, Steinwagner, Stern, Stone, Strada, Strafella, Stramaccioni, Surace, Sureau, Suyu, Swindells, Szafraniec, Szapudi, Taamoli, Talia, Tallada-Cresp\'{\i}, Tanidis, Tao, Tarr\'{\i}o, Tavagnacco, Taylor, Taylor, Taylor, Teixeira, Tenti, Teodoro~Idiago, Teplitz, Tereno, Tessore, Testa, Testera, Tewes, Teyssier, Theret, Thizy, Thomas, Toba, Toft, Toledo-Moreo, Tolstoy, Tommasi, Torbaniuk, Torradeflot, Tortora, Tosi, Tosti, Trifoglio, Troja, Trombetti, Tronconi, Tsedrik, Tsyganov, Tucci, Tutusaus, Uhlemann, Ulivi, Urbano, Vacher, Vaillon, Valdes, Valentijn, Valenziano, Valieri, Valiviita, Van~den Broeck, Vassallo, Vavrek, Venemans, Venhola, Ventura, Verdoes~Kleijn, Vergani, Verma, Vernizzi,
  Veropalumbo, Verza, Vescovi, Vibert, Viel, Vielzeuf, Viglione, Viitanen, Villaescusa-Navarro, Vinciguerra, Visticot, Voggel, von Wietersheim-Kramsta, Vriend, Wachter, Walmsley, Walth, Walton, Walton, Wander, Wang, Wang, Weaver, Weller, Whalen, Wiesmann, Wilde, Williams, Winther, Wittje, Wong, Wright, Yankelevich, Yeung, Youles, Yung, Zacchei, Zalesky, Zamorani, Zamorano~Vitorelli, Zanoni~Marc, Zennaro, Zerbi, Zinchenko, Zoubian, Zucca, \& Zumalacarregui}]{Mellier2024euclid}
Euclid Collaboration:~Mellier, Y., Abdurro'uf, Acevedo~Barroso, J.~A., {et~al.} 2024, arXiv e-prints, arXiv:2405.13491, \dodoi{10.48550/arXiv.2405.13491}

\bibitem[{Fenech~Conti {et~al.}(2017)Fenech~Conti, Herbonnet, Hoekstra, Merten, Miller, \& Viola}]{FenechConti2017}
Fenech~Conti, I., Herbonnet, R., Hoekstra, H., {et~al.} 2017, MNRAS, 467, 1627, \dodoi{10.1093/mnras/stx200}

\bibitem[{Goodfellow {et~al.}(2014)Goodfellow, Pouget-Abadie, Mirza, Xu, Warde-Farley, Ozair, Courville, \& Bengio}]{goodfellow2014GAN}
Goodfellow, I.~J., Pouget-Abadie, J., Mirza, M., {et~al.} 2014, Generative Adversarial Networks.
\newblock \doarXiv{1406.2661}

\bibitem[{{Grogin} {et~al.}(2011){Grogin}, {Kocevski}, {Faber}, {Ferguson}, {Koekemoer}, {Riess}, {Acquaviva}, {Alexander}, {Almaini}, {Ashby}, {Barden}, {Bell}, {Bournaud}, {Brown}, {Caputi}, {Casertano}, {Cassata}, {Castellano}, {Challis}, {Chary}, {Cheung}, {Cirasuolo}, {Conselice}, {Roshan Cooray}, {Croton}, {Daddi}, {Dahlen}, {Dav{\'e}}, {de Mello}, {Dekel}, {Dickinson}, {Dolch}, {Donley}, {Dunlop}, {Dutton}, {Elbaz}, {Fazio}, {Filippenko}, {Finkelstein}, {Fontana}, {Gardner}, {Garnavich}, {Gawiser}, {Giavalisco}, {Grazian}, {Guo}, {Hathi}, {H{\"a}ussler}, {Hopkins}, {Huang}, {Huang}, {Jha}, {Kartaltepe}, {Kirshner}, {Koo}, {Lai}, {Lee}, {Li}, {Lotz}, {Lucas}, {Madau}, {McCarthy}, {McGrath}, {McIntosh}, {McLure}, {Mobasher}, {Moustakas}, {Mozena}, {Nandra}, {Newman}, {Niemi}, {Noeske}, {Papovich}, {Pentericci}, {Pope}, {Primack}, {Rajan}, {Ravindranath}, {Reddy}, {Renzini}, {Rix}, {Robaina}, {Rodney}, {Rosario}, {Rosati}, {Salimbeni}, {Scarlata}, {Siana}, {Simard}, {Smidt}, {Somerville}, {Spinrad},
  {Straughn}, {Strolger}, {Telford}, {Teplitz}, {Trump}, {van der Wel}, {Villforth}, {Wechsler}, {Weiner}, {Wiklind}, {Wild}, {Wilson}, {Wuyts}, {Yan}, \& {Yun}}]{2011Grogin}
{Grogin}, N.~A., {Kocevski}, D.~D., {Faber}, S.~M., {et~al.} 2011, \apjs, 197, 35, \dodoi{10.1088/0067-0049/197/2/35}

\bibitem[{{Guo} {et~al.}(2013){Guo}, {Ferguson}, {Giavalisco}, {Barro}, {Willner}, {Ashby}, {Dahlen}, {Donley}, {Faber}, {Fontana}, {Galametz}, {Grazian}, {Huang}, {Kocevski}, {Koekemoer}, {Koo}, {McGrath}, {Peth}, {Salvato}, {Wuyts}, {Castellano}, {Cooray}, {Dickinson}, {Dunlop}, {Fazio}, {Gardner}, {Gawiser}, {Grogin}, {Hathi}, {Hsu}, {Lee}, {Lucas}, {Mobasher}, {Nandra}, {Newman}, \& {van der Wel}}]{Guo2013}
{Guo}, Y., {Ferguson}, H.~C., {Giavalisco}, M., {et~al.} 2013, \apjs, 207, 24, \dodoi{10.1088/0067-0049/207/2/24}

\bibitem[{{Hemmati} {et~al.}(2019){Hemmati}, {Capak}, {Masters}, {Davidzon}, {Dor{\`e}}, {Kruk}, {Mobasher}, {Rhodes}, {Scolnic}, \& {Stern}}]{Hemmati2019}
{Hemmati}, S., {Capak}, P., {Masters}, D., {et~al.} 2019, \apj, 877, 117, \dodoi{10.3847/1538-4357/ab1be5}

\bibitem[{Hemmati {et~al.}(2022)Hemmati, Huff, Nayyeri, Ferté, Melchior, Mobasher, Rhodes, Shahidi, \& Teplitz}]{Hemmati_2022}
Hemmati, S., Huff, E., Nayyeri, H., {et~al.} 2022, \apj, 941, 141, \dodoi{10.3847/1538-4357/aca1b8}

\bibitem[{{Hemmati} {et~al.}(2022){Hemmati}, {Huff}, {Nayyeri}, {Fert{\'e}}, {Melchior}, {Mobasher}, {Rhodes}, {Shahidi}, \& {Teplitz}}]{Hemmati2022}
{Hemmati}, S., {Huff}, E., {Nayyeri}, H., {et~al.} 2022, \apj, 941, 141, \dodoi{10.3847/1538-4357/aca1b8}

\bibitem[{Ho {et~al.}(2020)Ho, Jain, \& Abbeel}]{ho2020DDPM}
Ho, J., Jain, A., \& Abbeel, P. 2020, in Advances in Neural Information Processing Systems, ed. H.~Larochelle, M.~Ranzato, R.~Hadsell, M.~Balcan, \& H.~Lin, Vol.~33 (Curran Associates, Inc.), 6840--6851.
\newblock \url{https://proceedings.neurips.cc/paper_files/paper/2020/file/4c5bcfec8584af0d967f1ab10179ca4b-Paper.pdf}

\bibitem[{{Holzschuh} {et~al.}(2022){Holzschuh}, {O'Riordan}, {Vegetti}, {Rodriguez-Gomez}, \& {Thuerey}}]{Holzschuh2022}
{Holzschuh}, B.~J., {O'Riordan}, C.~M., {Vegetti}, S., {Rodriguez-Gomez}, V., \& {Thuerey}, N. 2022, \mnras, 515, 652, \dodoi{10.1093/mnras/stac1188}

\bibitem[{{Ivezi{\'c}} {et~al.}(2019){Ivezi{\'c}}, {Kahn}, {Tyson}, {Abel}, {Acosta}, {Allsman}, {Alonso}, {AlSayyad}, {Anderson}, {Andrew}, {Angel}, {Angeli}, {Ansari}, {Antilogus}, {Araujo}, {Armstrong}, {Arndt}, {Astier}, {Aubourg}, {Auza}, {Axelrod}, {Bard}, {Barr}, {Barrau}, {Bartlett}, {Bauer}, {Bauman}, {Baumont}, {Bechtol}, {Bechtol}, {Becker}, {Becla}, {Beldica}, {Bellavia}, {Bianco}, {Biswas}, {Blanc}, {Blazek}, {Blandford}, {Bloom}, {Bogart}, {Bond}, {Booth}, {Borgland}, {Borne}, {Bosch}, {Boutigny}, {Brackett}, {Bradshaw}, {Brandt}, {Brown}, {Bullock}, {Burchat}, {Burke}, {Cagnoli}, {Calabrese}, {Callahan}, {Callen}, {Carlin}, {Carlson}, {Chandrasekharan}, {Charles-Emerson}, {Chesley}, {Cheu}, {Chiang}, {Chiang}, {Chirino}, {Chow}, {Ciardi}, {Claver}, {Cohen-Tanugi}, {Cockrum}, {Coles}, {Connolly}, {Cook}, {Cooray}, {Covey}, {Cribbs}, {Cui}, {Cutri}, {Daly}, {Daniel}, {Daruich}, {Daubard}, {Daues}, {Dawson}, {Delgado}, {Dellapenna}, {de Peyster}, {de Val-Borro}, {Digel}, {Doherty}, {Dubois},
  {Dubois-Felsmann}, {Durech}, {Economou}, {Eifler}, {Eracleous}, {Emmons}, {Fausti Neto}, {Ferguson}, {Figueroa}, {Fisher-Levine}, {Focke}, {Foss}, {Frank}, {Freemon}, {Gangler}, {Gawiser}, {Geary}, {Gee}, {Geha}, {Gessner}, {Gibson}, {Gilmore}, {Glanzman}, {Glick}, {Goldina}, {Goldstein}, {Goodenow}, {Graham}, {Gressler}, {Gris}, {Guy}, {Guyonnet}, {Haller}, {Harris}, {Hascall}, {Haupt}, {Hernandez}, {Herrmann}, {Hileman}, {Hoblitt}, {Hodgson}, {Hogan}, {Howard}, {Huang}, {Huffer}, {Ingraham}, {Innes}, {Jacoby}, {Jain}, {Jammes}, {Jee}, {Jenness}, {Jernigan}, {Jevremovi{\'c}}, {Johns}, {Johnson}, {Johnson}, {Jones}, {Juramy-Gilles}, {Juri{\'c}}, {Kalirai}, {Kallivayalil}, {Kalmbach}, {Kantor}, {Karst}, {Kasliwal}, {Kelly}, {Kessler}, {Kinnison}, {Kirkby}, {Knox}, {Kotov}, {Krabbendam}, {Krughoff}, {Kub{\'a}nek}, {Kuczewski}, {Kulkarni}, {Ku}, {Kurita}, {Lage}, {Lambert}, {Lange}, {Langton}, {Le Guillou}, {Levine}, {Liang}, {Lim}, {Lintott}, {Long}, {Lopez}, {Lotz}, {Lupton}, {Lust}, {MacArthur}, {Mahabal},
  {Mandelbaum}, {Markiewicz}, {Marsh}, {Marshall}, {Marshall}, {May}, {McKercher}, {McQueen}, {Meyers}, {Migliore}, {Miller}, {Mills}, {Miraval}, {Moeyens}, {Moolekamp}, {Monet}, {Moniez}, {Monkewitz}, {Montgomery}, {Morrison}, {Mueller}, {Muller}, {Mu{\~n}oz Arancibia}, {Neill}, {Newbry}, {Nief}, {Nomerotski}, {Nordby}, {O'Connor}, {Oliver}, {Olivier}, {Olsen}, {O'Mullane}, {Ortiz}, {Osier}, {Owen}, {Pain}, {Palecek}, {Parejko}, {Parsons}, {Pease}, {Peterson}, {Peterson}, {Petravick}, {Libby Petrick}, {Petry}, {Pierfederici}, {Pietrowicz}, {Pike}, {Pinto}, {Plante}, {Plate}, {Plutchak}, {Price}, {Prouza}, {Radeka}, {Rajagopal}, {Rasmussen}, {Regnault}, {Reil}, {Reiss}, {Reuter}, {Ridgway}, {Riot}, {Ritz}, {Robinson}, {Roby}, {Roodman}, {Rosing}, {Roucelle}, {Rumore}, {Russo}, {Saha}, {Sassolas}, {Schalk}, {Schellart}, {Schindler}, {Schmidt}, {Schneider}, {Schneider}, {Schoening}, {Schumacher}, {Schwamb}, {Sebag}, {Selvy}, {Sembroski}, {Seppala}, {Serio}, {Serrano}, {Shaw}, {Shipsey}, {Sick}, {Silvestri},
  {Slater}, {Smith}, {Smith}, {Sobhani}, {Soldahl}, {Storrie-Lombardi}, {Stover}, {Strauss}, {Street}, {Stubbs}, {Sullivan}, {Sweeney}, {Swinbank}, {Szalay}, {Takacs}, {Tether}, {Thaler}, {Thayer}, {Thomas}, {Thornton}, {Thukral}, {Tice}, {Trilling}, {Turri}, {Van Berg}, {Vanden Berk}, {Vetter}, {Virieux}, {Vucina}, {Wahl}, {Walkowicz}, {Walsh}, {Walter}, {Wang}, {Wang}, {Warner}, {Wiecha}, {Willman}, {Winters}, {Wittman}, {Wolff}, {Wood-Vasey}, {Wu}, {Xin}, {Yoachim}, \& {Zhan}}]{Ivezic2019LSST}
{Ivezi{\'c}}, {\v{Z}}., {Kahn}, S.~M., {Tyson}, J.~A., {et~al.} 2019, \apj, 873, 111, \dodoi{10.3847/1538-4357/ab042c}

\bibitem[{{Kannawadi} {et~al.}(2019){Kannawadi}, {Hoekstra, Henk}, {Miller, Lance}, {Viola, Massimo}, {Fenech Conti, Ian}, {Herbonnet, Ricardo}, {Erben, Thomas}, {Heymans, Catherine}, {Hildebrandt, Hendrik}, {Kuijken, Konrad}, {Vakili, Mohammadjavad}, \& {Wright, Angus H.}}]{Kannawadi2019}
{Kannawadi}, A., {Hoekstra, Henk}, {Miller, Lance}, {et~al.} 2019, A\&A, 624, A92, \dodoi{10.1051/0004-6361/201834819}

\bibitem[{Karras {et~al.}(2022)Karras, Aittala, Aila, \& Laine}]{karras2022elucidating}
Karras, T., Aittala, M., Aila, T., \& Laine, S. 2022, Advances in neural information processing systems, 35, 26565

\bibitem[{{Kingma} \& {Welling}(2013)}]{Kingma2013}
{Kingma}, D.~P., \& {Welling}, M. 2013, arXiv e-prints, arXiv:1312.6114, \dodoi{10.48550/arXiv.1312.6114}

\bibitem[{{Koekemoer} {et~al.}(2011){Koekemoer}, {Faber}, {Ferguson}, {Grogin}, {Kocevski}, {Koo}, {Lai}, {Lotz}, {Lucas}, {McGrath}, {Ogaz}, {Rajan}, {Riess}, {Rodney}, {Strolger}, {Casertano}, {Castellano}, {Dahlen}, {Dickinson}, {Dolch}, {Fontana}, {Giavalisco}, {Grazian}, {Guo}, {Hathi}, {Huang}, {van der Wel}, {Yan}, {Acquaviva}, {Alexander}, {Almaini}, {Ashby}, {Barden}, {Bell}, {Bournaud}, {Brown}, {Caputi}, {Cassata}, {Challis}, {Chary}, {Cheung}, {Cirasuolo}, {Conselice}, {Roshan Cooray}, {Croton}, {Daddi}, {Dav{\'e}}, {de Mello}, {de Ravel}, {Dekel}, {Donley}, {Dunlop}, {Dutton}, {Elbaz}, {Fazio}, {Filippenko}, {Finkelstein}, {Frazer}, {Gardner}, {Garnavich}, {Gawiser}, {Gruetzbauch}, {Hartley}, {H{\"a}ussler}, {Herrington}, {Hopkins}, {Huang}, {Jha}, {Johnson}, {Kartaltepe}, {Khostovan}, {Kirshner}, {Lani}, {Lee}, {Li}, {Madau}, {McCarthy}, {McIntosh}, {McLure}, {McPartland}, {Mobasher}, {Moreira}, {Mortlock}, {Moustakas}, {Mozena}, {Nandra}, {Newman}, {Nielsen}, {Niemi}, {Noeske}, {Papovich},
  {Pentericci}, {Pope}, {Primack}, {Ravindranath}, {Reddy}, {Renzini}, {Rix}, {Robaina}, {Rosario}, {Rosati}, {Salimbeni}, {Scarlata}, {Siana}, {Simard}, {Smidt}, {Snyder}, {Somerville}, {Spinrad}, {Straughn}, {Telford}, {Teplitz}, {Trump}, {Vargas}, {Villforth}, {Wagner}, {Wandro}, {Wechsler}, {Weiner}, {Wiklind}, {Wild}, {Wilson}, {Wuyts}, \& {Yun}}]{Koekemoer2011}
{Koekemoer}, A.~M., {Faber}, S.~M., {Ferguson}, H.~C., {et~al.} 2011, \apjs, 197, 36, \dodoi{10.1088/0067-0049/197/2/36}

\bibitem[{Lanusse {et~al.}(2021)Lanusse, Mandelbaum, Ravanbakhsh, Li, Freeman, \& Póczos}]{Lanusse2021}
Lanusse, F., Mandelbaum, R., Ravanbakhsh, S., {et~al.} 2021, MNRAS, 504, 5543, \dodoi{10.1093/mnras/stab1214}

\bibitem[{Laureijs {et~al.}(2011)Laureijs, Amiaux, Arduini, Auguères, Brinchmann, Cole, Cropper, Dabin, Duvet, Ealet, Garilli, Gondoin, Guzzo, Hoar, Hoekstra, Holmes, Kitching, Maciaszek, Mellier, Pasian, Percival, Rhodes, Criado, Sauvage, Scaramella, Valenziano, Warren, Bender, Castander, Cimatti, Fèvre, Kurki-Suonio, Levi, Lilje, Meylan, Nichol, Pedersen, Popa, Lopez, Rix, Rottgering, Zeilinger, Grupp, Hudelot, Massey, Meneghetti, Miller, Paltani, Paulin-Henriksson, Pires, Saxton, Schrabback, Seidel, Walsh, Aghanim, Amendola, Bartlett, Baccigalupi, Beaulieu, Benabed, Cuby, Elbaz, Fosalba, Gavazzi, Helmi, Hook, Irwin, Kneib, Kunz, Mannucci, Moscardini, Tao, Teyssier, Weller, Zamorani, Osorio, Boulade, Foumond, Giorgio, Guttridge, James, Kemp, Martignac, Spencer, Walton, Blümchen, Bonoli, Bortoletto, Cerna, Corcione, Fabron, Jahnke, Ligori, Madrid, Martin, Morgante, Pamplona, Prieto, Riva, Toledo, Trifoglio, Zerbi, Abdalla, Douspis, Grenet, Borgani, Bouwens, Courbin, Delouis, Dubath, Fontana, Frailis,
  Grazian, Koppenhöfer, Mansutti, Melchior, Mignoli, Mohr, Neissner, Noddle, Poncet, Scodeggio, Serrano, Shane, Starck, Surace, Taylor, Verdoes-Kleijn, Vuerli, Williams, Zacchei, Altieri, Sanz, Kohley, Oosterbroek, Astier, Bacon, Bardelli, Baugh, Bellagamba, Benoist, Bianchi, Biviano, Branchini, Carbone, Cardone, Clements, Colombi, Conselice, Cresci, Deacon, Dunlop, Fedeli, Fontanot, Franzetti, Giocoli, Garcia-Bellido, Gow, Heavens, Hewett, Heymans, Holland, Huang, Ilbert, Joachimi, Jennins, Kerins, Kiessling, Kirk, Kotak, Krause, Lahav, van Leeuwen, Lesgourgues, Lombardi, Magliocchetti, Maguire, Majerotto, Maoli, Marulli, Maurogordato, McCracken, McLure, Melchiorri, Merson, Moresco, Nonino, Norberg, Peacock, Pello, Penny, Pettorino, Porto, Pozzetti, Quercellini, Radovich, Rassat, Roche, Ronayette, Rossetti, Sartoris, Schneider, Semboloni, Serjeant, Simpson, Skordis, Smadja, Smartt, Spano, Spiro, Sullivan, Tilquin, Trotta, Verde, Wang, Williger, Zhao, Zoubian, \& Zucca}]{laureijs2011euclid}
Laureijs, R., Amiaux, J., Arduini, S., {et~al.} 2011, arXiv:1110.3193.
\newblock \doarXiv{1110.3193}

\bibitem[{Lizarraga {et~al.}(2024)Lizarraga, Jiang, Nowack, Li, Wu, Boscoe, \& Do}]{lizarraga2024}
Lizarraga, A., Jiang, E.~H., Nowack, J., {et~al.} 2024, arXiv preprint arXiv:2411.18440

\bibitem[{Lu {et~al.}(2022)Lu, Zhou, Bao, Chen, Li, \& Zhu}]{lu2022dpm}
Lu, C., Zhou, Y., Bao, F., {et~al.} 2022, arXiv preprint arXiv:2211.01095

\bibitem[{{MacCrann} {et~al.}(2022){MacCrann}, {Becker}, {McCullough}, {Amon}, {Gruen}, {Jarvis}, {Choi}, {Troxel}, {Sheldon}, {Yanny}, {Herner}, {Dodelson}, {Zuntz}, {Eckert}, {Rollins}, {Varga}, {Bernstein}, {Gruendl}, {Harrison}, {Hartley}, {Sevilla-Noarbe}, {Pieres}, {Bridle}, {Myles}, {Alarcon}, {Everett}, {S{\'a}nchez}, {Huff}, {Tarsitano}, {Gatti}, {Secco}, {Abbott}, {Aguena}, {Allam}, {Annis}, {Bacon}, {Bertin}, {Brooks}, {Burke}, {Carnero Rosell}, {Carrasco Kind}, {Carretero}, {Costanzi}, {Crocce}, {Pereira}, {De Vicente}, {Desai}, {Diehl}, {Dietrich}, {Doel}, {Eifler}, {Ferrero}, {Fert{\'e}}, {Flaugher}, {Fosalba}, {Frieman}, {Garc{\'{\i}}a-Bellido}, {Gaztanaga}, {Gerdes}, {Giannantonio}, {Gschwend}, {Gutierrez}, {Hinton}, {Hollowood}, {Honscheid}, {James}, {Lahav}, {Lima}, {Maia}, {March}, {Marshall}, {Martini}, {Melchior}, {Menanteau}, {Miquel}, {Mohr}, {Morgan}, {Muir}, {Ogando}, {Palmese}, {Paz-Chinch{\'o}n}, {Plazas}, {Rodriguez-Monroy}, {Roodman}, {Samuroff}, {Sanchez}, {Scarpine}, {Serrano},
  {Smith}, {Soares-Santos}, {Suchyta}, {Swanson}, {Tarle}, {Thomas}, {To}, {Wilkinson}, {Wilkinson}, \& {DES Collaboration}}]{MacCrann2022}
{MacCrann}, N., {Becker}, M.~R., {McCullough}, J., {et~al.} 2022, \mnras, 509, 3371, \dodoi{10.1093/mnras/stab2870}

\bibitem[{Mandelbaum {et~al.}(2018)Mandelbaum, Lanusse, Leauthaud, Armstrong, Simet, Miyatake, Meyers, Bosch, Murata, Miyazaki, \& Tanaka}]{Mandelbaum_2018}
Mandelbaum, R., Lanusse, F., Leauthaud, A., {et~al.} 2018, MNRAS, 481, 3170–3195, \dodoi{10.1093/mnras/sty2420}

\bibitem[{Ronneberger {et~al.}(2015)Ronneberger, Fischer, \& Brox}]{ronneberger2015u}
Ronneberger, O., Fischer, P., \& Brox, T. 2015, in Medical image computing and computer-assisted intervention--MICCAI 2015: 18th international conference, Munich, Germany, October 5-9, 2015, proceedings, part III 18, Springer, 234--241

\bibitem[{{Rowe} {et~al.}(2015){Rowe}, {Jarvis}, {Mandelbaum}, {Bernstein}, {Bosch}, {Simet}, {Meyers}, {Kacprzak}, {Nakajima}, {Zuntz}, {Miyatake}, {Dietrich}, {Armstrong}, {Melchior}, \& {Gill}}]{Rowe2015}
{Rowe}, B.~T.~P., {Jarvis}, M., {Mandelbaum}, R., {et~al.} 2015, Astronomy and Computing, 10, 121, \dodoi{10.1016/j.ascom.2015.02.002}

\bibitem[{Smith {et~al.}(2022)Smith, Geach, Jackson, Arora, Stone, \& Courteau}]{Smith2022}
Smith, M.~J., Geach, J.~E., Jackson, R.~A., {et~al.} 2022, Monthly Notices of the Royal Astronomical Society, 511, 1808, \dodoi{10.1093/mnras/stac130}

\bibitem[{Song {et~al.}(2021)Song, Meng, \& Ermon}]{song2021denoising}
Song, J., Meng, C., \& Ermon, S. 2021, in International Conference on Learning Representations.
\newblock \url{https://openreview.net/forum?id=St1giarCHLP}

\bibitem[{Spergel {et~al.}(2015)Spergel, Gehrels, Baltay, Bennett, Breckinridge, Donahue, Dressler, Gaudi, Greene, Guyon, Hirata, Kalirai, Kasdin, Macintosh, Moos, Perlmutter, Postman, Rauscher, Rhodes, Wang, Weinberg, Benford, Hudson, Jeong, Mellier, Traub, Yamada, Capak, Colbert, Masters, Penny, Savransky, Stern, Zimmerman, Barry, Bartusek, Carpenter, Cheng, Content, Dekens, Demers, Grady, Jackson, Kuan, Kruk, Melton, Nemati, Parvin, Poberezhskiy, Peddie, Ruffa, Wallace, Whipple, Wollack, \& Zhao}]{spergel2015widefield}
Spergel, D., Gehrels, N., Baltay, C., {et~al.} 2015, Wide-Field InfrarRed Survey Telescope-Astrophysics Focused Telescope Assets WFIRST-AFTA 2015 Report.
\newblock \doarXiv{1503.03757}

\bibitem[{Spindler {et~al.}(2020)Spindler, Geach, \& Smith}]{Spindler_2020}
Spindler, A., Geach, J.~E., \& Smith, M.~J. 2020, MNRAS, 502, 985–1007, \dodoi{10.1093/mnras/staa3670}

\end{thebibliography}
\bibliographystyle{aasjournal}


\end{document}